\documentclass[preprint,aps,10pt]{revtex4} 
\usepackage{amssymb,latexsym,amsmath,cancel,graphicx,epstopdf}

\def\nn{\nonumber}
\def\wt{\tilde}
\def\ov{\overline}
\def\tnp{{\wt n}_+^p}
\def\tnm{{\wt n}_-^p}
\def\De{\Delta}
\def\de{\delta}
\def\gm{\gamma}
\def\Gm{\Gamma}
\def\Lm{\Lambda}

\def\Sg{\Sigma}

\def\mn{{\mu\nu}}
\def\om{\omega}
\def\tht{\theta}
\def\del{\partial}
\def\grad{\nabla}
\def\be{\begin{equation}}
\def\ee{\end{equation}}
\def\ba{\begin{eqnarray}}
\def\ea{\end{eqnarray}}
\def\m{\boldsymbol}

\begin{document}
\title{$\Delta$ self-energy at finite temperature and density and the $\pi N$ cross-section}
\author{Snigdha Ghosh}
\affiliation{Variable Energy Cyclotron Centre, 1/AF Bidhannagar, Kolkata - 700064, INDIA}
\affiliation{Homi Bhabha National Institute, Training School Complex, Anushaktinagar, Mumbai - 400085, INDIA}
\author{Sourav Sarkar}
\affiliation{Variable Energy Cyclotron Centre, 1/AF Bidhannagar, Kolkata - 700064, INDIA}
\affiliation{Homi Bhabha National Institute, Training School Complex, Anushaktinagar, Mumbai - 400085, INDIA}
\author{Sukanya Mitra}
\affiliation{Indian Institute of Technology Gandhinagar,  Gandhinagar-382355, Gujarat, INDIA}

%\date{}

\begin{abstract}
The self energy of $\Delta$-baryon is evaluated at finite temperature and density using the real time formalism of thermal field theory. The Dyson-Schwinger equation is used to get the exact thermal propagator followed by the spectral function of $\Delta$. The $\pi N$ scattering cross section obtained using explicit $\De$ exchange is normalized to the experimental data in vacuum and its medium modification is implemented by means of the exact thermal propagator. A significant suppression of the peak of the cross-section is observed at higher temperature and baryon density. Effects on the mean relaxation time of nucleons and the temperature dependence of the shear viscosity of a pion nucleon gas are demonstrated.
\end{abstract}
\maketitle
\section{Introduction}

The properties of hadrons under conditions of high temperature and/or baryon density have been widely studied owing to the possibility of restoration of chiral symmetry of QCD spontaneously broken by the vacuum~\cite{Bazavov}. Relativistic collisions of heavy ions provide the unique opportunity to actually produce hot/dense matter under controlled conditions and provide experimental verification of such studies (see e.g.~\cite{Rapp}). The spectral changes of hadrons as
a result of compression and/or heating thus play a very significant role in unraveling the dynamics of the underlying colour degrees of freedom and the vacuum structure of QCD~\cite{CBMBook}. The broadening of vector spectral function in the medium observed through the invariant mass spectra of dileptons in heavy ion collisions at SPS~\cite{Arnaldi} and RHIC~\cite{Adare} have been interpreted (see e.g.~\cite{Sarkar}) to have non-trivial, though indirect implications on chiral symmetry restoration.  In this connection the study of spectral properties of baryon resonances such as $\De$ and $N^*$ etc. are important; though not measurable directly because of final state interactions, they provide valuable input to the evaluation of the self-energies of low mass vector mesons which can be measured in heavy ion collision experiments through electromagnetic probes. In addition to broad resonances stable hadrons such as the nucleon also develop considerable widths in the medium~\cite{Ghosh2} and this is mainly due to resonant scattering with pions involving baryon resonances.

The in-medium self-energy of the $\De$ has played a key role in the understanding of dynamics of nuclear interactions particularly in the resonance region where it is excited as a $\pi N$ resonance~\cite{Ericson}. The propagation of $\pi$, $N$ and $\De$ thus become intimately connected. The description of pion-nucleus interaction for example, depends on the formation, propagation and decay of the $\De$ in the nuclear environment~\cite{Oset,Malfliet}. Models based on this picture, known as $\De$-hole models, successfully explain many scattering phenomena in elastic as well as inelastic channels. The $\De$ self-energy extracted using this approach has also found applications in transport models describing the dynamical evolution of nucleus-nucleus collisions~\cite{Mosel,Wolf}.

%Study of transport properties of hot/dense hadronic matter which is produced towards the later stages of relativistic heavy ion collisions provides a suitable %arena to probe the behavior of scattering amplitudes in the medium. 
The  spectral modification of the $\De$ at finite temperature and density and the in-medium $\pi N$ cross-section remains an interesting issue of discussion. A very useful and topical application of this study is to investigate the medium dependence of the shear viscosity of hot and dense hadronic matter which is produced towards the later stages of relativistic heavy ion collisions.
In the kinetic theory approach of evaluating transport coefficients like shear and bulk viscosities, thermal conductivity etc. the cross-section enters as the dynamical input. Since pions account for most of the multiplicity in relativistic heavy ion collisions, the case of the pion gas has received some attention in recent times. Scattering amplitudes evaluated using chiral perturbation theory to lowest order have been used in~\cite{Santalla,Chen} and unitarization improved estimates were employed in~\cite{Dobado} to evaluate the shear viscosity. Again, phenomenological scattering cross-section using experimental phase shifts were utilized in~\cite{Prakash,Chen,Davesne,Itakura} to obtain the viscous coefficients.
While in~\cite{Moore,Juan} the effect of number changing processes on the bulk viscosity of a pion gas has been studied, in~\cite{Fraile} unitarized chiral perturbation theory was used to demonstrate the breaking of conformal symmetry by the pion mass. It is important to point out that in all these approaches vacuum amplitudes have been used. However, cross-sections of scattering between excitations in the medium could be non-trivially affected by the presence of other particles constituting it. It follows that in addition to the usual quantum fluctuations, thermal fluctuations modify the propagation of mediating particles and consequently the invariant amplitude. The case of the $\pi\pi$ cross-section in a hot pion gas was studied by Barz et al~\cite{Barz} obtaining a substantial reduction of the magnitude in the region of the $\rho$ peak. With this motivating feature, recently~\cite{Mitra1}, the amplitudes for elastic $\pi\pi$ scattering were evaluated using in-medium propagators for the exchanged $\rho$ and $\sigma$ mesons incorporating loop graphs with $\pi$, $\omega$, $h_1$ and $a_1$ mesons in the internal lines~\cite{Ghosh1}. The cross-section so obtained showed a substantial reduction in magnitude at the peak position
at higher values of temperature. When used in the collision integral of the transport equation, these medium modified scattering amplitudes were found to significantly affect the temperature dependence of the viscosities~\cite{Mitra2} and thermal conductivity~\cite{Mitra3} of a hot pion gas.

Based on our experience with the pion gas and keeping in view the upcoming CBM experiment at FAIR it is  natural to ask how the presence of a finite baryon density in addition to temperature is likely to affect the shear viscosity. To study such effects one has to include nucleons and consequently the $\pi N$ cross-section becomes the principal dynamical factor in its evaluation. Incorporating the in-medium $\pi N$ cross-section calculated using the modified $\De$ self-energy a more reliable estimate of the shear viscosity, in particular it's dependence on temperature and baryon density can be obtained. When used as input in the viscous hydrodynamic equations a more realistic scenario of space time evolution of the later stages of heavy ion collisions is likely to be achieved.

In this work we obtain the $\De$ self-energy at finite temperature and baryon density evaluating several one-loop diagrams with $\pi$, $\rho$, $N$ and $\De$ in the internal lines using standard thermal field theoretic methods. The in-medium propagator of the $\De$ is then used in the scattering amplitudes to obtain the $\pi N$ cross-section. This is utilized to evaluate the relaxation times in a hadronic gas mixture of pions and nucleons. Finally, the temperature and density dependence of the shear viscosity is obtained.

Medium modifications of the $\De$ resonance has been studied mostly in nuclear matter~\cite{Oset,Wehrberger,Korpa1,Kim,Xia}. Whereas a many body expansion in terms of particle-hole excitations has been used in~\cite{Oset} to evaluate the $\De$ self-energy in nuclear matter, in~\cite{Wehrberger} and~\cite{Kim} its decay in the medium is investigated using the quantum hadrodynamics (QHD) model. 
A self-consistent treatment of pions and $\De$'s  in nuclear matter at zero and finite temperature may be found in~\cite{Xia} and~\cite{Korpa2} respectively. In~\cite{Korpa1}, the $\De$ self-energy due to $\pi N$ loop in nuclear matter is obtained in a relativistic approach. More recently, modification of the $\De$ spectral function at finite temperature and density due to resonant scattering off thermal pions has been obtained in~\cite{vanHees}.
In addition to these theoretical studies, properties of the $\De$ have been studied experimentally using invariant mass analyses of $\pi N$ pairs~\cite{Hjort,Mishra}.

In the next section we recall some basic features of the real-time version of thermal field theory. Then we evaluate the $\De$ self-energy and discuss numerical results. In the subsequent section we evaluate the amplitudes of $\pi N$ scattering leading to the cross-section in the medium. This is followed by a section on the shear viscosity of a $\pi N$ gas and finally by summary and discussions. Some details of the calculation is provided in the Appendix.

\section{The $\De$ self-energy in the medium}
\subsection{The in-medium propagators in the real time formalism}

In the real time formalism of thermal field theory, all two-point functions including the self-energy take the form of $2\times2$ matrices~\cite{Mallik,Bellac}. But each of the matrices may be diagonalized, when it is given essentially by a single analytic function which determines completely the dynamics of the corresponding two-point function. This function being given by any one, say the 11-component of the matrix, we need to evaluate only this component of the self-energy matrix. In the following we specify only the 11-component of the thermal propagator for the particles involved in the one-loop graphs that are considered in this work.

The 11-component of a free thermal propagator matrix for a particle consists of its vacuum propagator and a term depending on the on-shell distribution function of like particles in the medium through which it propagates. The form of the latter term depends only on whether the particle in question is a boson or a fermion. 

The 11-component of the thermal pion propagator is given by
\be 
D_{11}(k,m_\pi)=\De(k,m_\pi) +2\pi iN_1^2(k,m_\pi)\delta(k^2-m_\pi^2)
\ee
where
\ba
\De(k,m) &=& \frac{-1}{k^2-m^2+i\eta} \nn\\
N_1(k,m)&=&\theta(k^0)\sqrt{n_+^k} + \theta(-k^0)\sqrt{n_-^k}
\ea
with
\[n_\pm^k=\frac{1}{e^{\beta\left(\omega_k\mp\mu_k\right)}-1}~~;~~\omega_k=\sqrt{\vec{k}^2+m^2}\]
and $\theta(k^0)=1$ for $k^0>0$ and 0 for $k^0<0$.

The thermal part remaining same, the $\rho$ propagator with the familiar polarization sum follows as 
\be 
D_{11}^\mn(k,m_\rho)=\left(-g^\mn+\frac{k^\mu k^\nu}{m_\rho^2}\right)D_{11}(k,m_\rho)~.
\ee

We now consider the fermionic propagators whose thermal (matrix) parts are different from the bosonic ones discussed above.
The 11-components of the nucleon and $\De$ propagators are respectively given by
\be 
S^{11}(p)=(\cancel p+m_N)E^{11}(p,m_N)
\ee
and
\ba 
S^{11}_\mn(p)=(\cancel p+m_\De)&&\left\{-g_\mn+\frac{2}{3m_\De^2}p_\mu p_\nu+\frac{1}{3}\gm_\mu \gm_\nu\right.\nn\\
&&\left.+\frac{1}{3m_\De}(\gm_\mu p_\nu-\gm_\nu p_\mu)\right\}E^{11}(p,m_\De)
\ea
where, $E_{11}(p,m)$ is given in Appendix-A.

The complete propagator $\m{S}'$ is given by the Dyson equation in terms of the free fermion propagator $\m{S}$ and self-energy $\m\Pi$,
\be
\m{S}'=\m{S}-\m{S}\, \m{\Pi}\, \m{S}'
\label{dyson-schwinger}
\ee
where each is a 2$\times$2 matrix in the thermal indices. They can be diagonalized to get the respective analytic functions, denoted by bar, so that
\be 
\ov{S}'=\ov{S}-\ov{S}\,{\ov\Pi}\,\ov{S}'~.
\label{dyson-scalar}
\ee
%Recalling that $\ov D=\De$ this equation is easily solved to give
%\begin{equation}
%\ov D'=\frac{-1}{k^2-m^2+\ov{\Pi}}~.
%\end{equation}
The self-energy function $\ov{\Pi}$ can be obtained from any single component of the self energy matrix as discussed in Appendix-A. It is related to, say, the 11-component by 
\ba
{\rm Im}\bar{\Pi}(p)&=&\epsilon(p_0)\coth[\beta (p_0-\mu_p)/2] {\rm Im}\Pi_{11}(p)\nn\\
{\rm Re}\bar{\Pi}(p)&=&{\rm Re}\Pi_{11}(p)
\label{real-imag}
\ea
where $\epsilon(p_0)=+1$ for $p_0>0$ and $-1$ for $p_0<0$.

\subsection{The $\Delta$ self-energy}

\begin{figure}
\centering
\includegraphics[scale=0.5,angle=0]{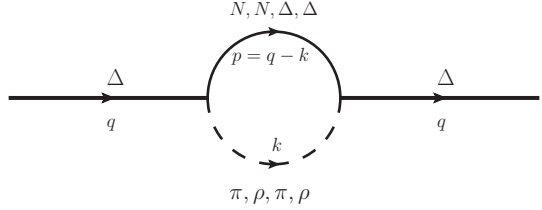}
\caption{Feynman Diagrams for $\Delta$-Self Energy}
\label{fig:Delta_SelfEnergy}
\end{figure}

We begin by writing down the expressions for the one-loop self-energy graphs for the  $\De$ {\it in vacuum}. 
For the four cases shown in fig.~\ref{fig:Delta_SelfEnergy} they can be generally expressed as 
\begin{equation}
\Pi^{\mu\nu}_{vac}(q)=i\int\frac{d^4k}{(2\pi)^4}N^{\mu\nu}\De(p)\De(k)
\end{equation}
where $N^{\mu\nu}$ contains terms coming from the two vertices and the spin factors appearing in the propagators for the internal lines. They can be read off from the expressions for $\Pi^\mn$ given in Appendix-B and appear as 
\begin{eqnarray}
N_{\pi N\Delta}^{\mu\nu}&=&\frac{f_{\pi N\Delta}^2}{m_\pi^2}F^2(p,k)\left[k_\alpha k_\beta\mathcal{O}^{\nu\beta}(\cancel{p}+m_p)\mathcal{O}^{\alpha\mu}\right]\\
N_{\rho N\Delta}^{\mu\nu}&=&\frac{f_{\rho N\Delta}^2}{m_\rho^2}F^2(p,k) \left[ \mathcal{O}^{\nu\eta}\gamma^5 \left( \gamma_\beta k_\eta-g_{\beta\eta}\cancel{k}\right)\left(\cancel{p}+m_p\right)\gamma^5\left( \gamma_\alpha k_\sigma-g_{\alpha\sigma}\cancel{k}\right)\mathcal{O}^{\mu\sigma}A^{\alpha\beta} \right]\\
N_{\pi\Delta\Delta}^{\mu\nu}&=&\frac{f_{\pi\Delta\Delta}^2}{m_\pi^2}F^2(p,k) \left[g_{\chi\psi}g_{\eta\phi}\mathcal{O}^{\nu\chi}\gamma^5\cancel{k}\mathcal{O}^{\psi\sigma}\Sigma_{\lambda\sigma}(p)\mathcal{O}^{\lambda\eta}\gamma^5\cancel{k}\mathcal{O}^{\phi\mu}\right]\\
N_{\rho\Delta\Delta}^{\mu\nu}&=&f_{\rho\Delta\Delta}^2 F^2(p,k) \left[ g_{\chi\psi}g_{\eta\phi}\mathcal{O}^{\nu\chi} \left( \gamma^\beta + i\frac{\kappa_{\rho\Delta\Delta}}{2m_\Delta}\sigma^{\beta\epsilon}k_\epsilon \right) \mathcal{O}^{\psi\theta}\Sigma_{\lambda\theta}(p)\right.\nn\\&&\hskip 3cm\left.\mathcal{O}^{\lambda\eta} \left( \gamma^\alpha - i\frac{\kappa_{\rho\Delta\Delta}}{2m_\Delta}\sigma^{\alpha\delta}k_\delta \right) \mathcal{O}^{\phi\mu}A_{\alpha\beta} \right]
\label{nmunu}
\end{eqnarray}
where
\begin{eqnarray}
A_{\alpha\beta}(k)&=&-g_{\alpha\beta}+\frac{k_\alpha k_\beta}{m_k^2}~~~\text{and}\nn\\
\Sigma_{\alpha\beta}(q)&=&\left(\cancel{q}+m_q\right)\left[ -g_{\alpha\beta}+\frac{1}{3m_q^2}q_\alpha q_\beta + \frac{1}{3}\gamma_\alpha\gamma_\beta+\frac{1}{3m_q}\left( \gamma_\alpha q_\beta-\gamma_\beta q_\alpha \right) \right]~.
\label{projectors}
\end{eqnarray}

We now proceed to write down the corresponding expressions {\em in the medium}.
As discussed above, we need to evaluate only the 11-component $\Pi^{\mu\nu}_{11}$ which is obtained 
by replacing the vacuum propagators by the 11-component of thermal propagators given in the previous section. %Realizing that the spin sums remain unaffected 
The self-energy in the medium is thus given by
\begin{equation}
\Pi^{\mu\nu}_{11}(q)=i\int\frac{d^4k}{(2\pi)^4}N^{\mu\nu}E_{11}(p)D_{11}(k)~.
\label{eqn_self_energy}
\end{equation}
 
Expanding $D_{11}$ and $E_{11}$ we obtain in addition to $\Pi^\mn_{vac}$, two terms which are linear in the thermal distribution function and the fourth term non-linear in the distribution function which is purely imaginary. 
Performing the $k^0$ integral and using eqns.~(\ref{real-imag}) we obtain the imaginary and real parts
of the self-energy function as
\begin{eqnarray}
\text{Im}\bar{\Pi}^{\mu\nu}(q)&=& -\pi\epsilon(q_0)\int\frac{d^3k}{(2\pi)^3}\frac{1}{4\omega_k\omega_p}\times\nn\\
&&[N^{\mu\nu}(k^0=\omega_k)\{(1+n_+^k-\tilde{n}_+^p)\delta(q_0-\omega_k-\omega_p)+(-n_+^k-\tilde{n}_-^p)
\delta(q_0-\omega_k+\omega_p)\}+\nn\\ 
&&N^{\mu\nu}(k^0=-\omega_k)\{(-1-n_-^k+\tilde{n}_-^p)\delta(q_0+\omega_k+\omega_p)+
(n_-^k+\tilde{n}_+^p)\delta(q_0+\omega_k-\omega_p)\}]\nn\\
\label{im1}
\end{eqnarray}
and
\begin{eqnarray}
\text{Re}\bar{\Pi}^{\mu\nu}(q) = \int\frac{d^3k}{(2\pi)^3}\frac{1}{2\omega_k\omega_p}&&\mathcal{P}\left[
\left(\frac{n_+^k\omega_p N^{\mu\nu}(k^0=\omega_k)}{(q_0-\omega_k)^2-\omega_p^2}\right) +
\left(\frac{n_-^k\omega_p N^{\mu\nu}(k^0=-\omega_k)}{(q_0+\omega_k)^2-\omega_p^2}\right)-\right. \nn\\ 
&&\left.\left(\frac{\tilde{n}_+^p\omega_k N^{\mu\nu}(k^0=q_0-\omega_p)}{(q_0-\omega_p)^2-\omega_k^2}\right) -
\left(\frac{\tilde{n}_-^p\omega_k N^{\mu\nu}(k^0=q_0+\omega_p)}{(q_0+\omega_p)^2-\omega_k^2}\right) \right]\nn\\
\end{eqnarray}
where $\om_k=\sqrt{m_k^2+\vec k^2}$ and $\om_p=\sqrt{m_p^2+(\vec q-\vec k)^2}$.

Each of the terms in the imaginary part can be related to scattering and decay of the $\De$-baryon. The delta functions in the four terms define the kinematic domains where these processes occur.
The regions where these are non-vanishing correspond to branch cuts in the complex $q_0$ plane. The first and third terms in (\ref{im1}) are non-vanishing for $q^2>(m_k+m_p)^2$. This is the usual unitary cut already present in vacuum. The second and fourth terms which are non-zero for $q^2<(m_p-m_k)^2$ defines the Landau cut and is purely a medium effect.
Confining ourselves to the kinematic region $q_0>0$ and $q^2>0$, the first and fourth terms 
only contribute. The first corresponds to absorption of the $\De$ due to decay into a baryon-meson pair  such as $N\pi$, $N\rho$ etc. and is thus weighted by a thermal factor $1+n_+^k-\tilde{n}_+^p=(1+n_+^k)(1-\tilde{n}_+^p)+n_+^k \tilde{n}_+^p$, indicating Bose enhancement of the meson and Pauli blocking of the baryon in the process $\De\to\pi N$ plus the  usual thermal factors for the initial state in the time reversed process where the $\De$ is produced. The fourth term is due to absorption of the $\De$ due to scattering from a meson producing a baryon in the final state and vice versa as is evident from the thermal weight factor $n_-^k+\tilde{n}_+^p=n_-^k(1-\tilde{n}_+^p)+\tilde{n}_+^p(1+n_-^k)$.

To take into account the finite width of unstable particles in the loop graphs the self-energy is folded with their spectral functions. As a consequence the sharp thresholds of the branch cuts get smeared. 
For unstable mesons ($h$) we use~\cite{Gonzalez}, 
\be
\Pi(q,m_h)= \frac{1}{N_h}\int^{(m_h+2\Gm_h)^2}_{(m_h-2\Gm_h)^2}dM^2\frac{1}{\pi} 
{\rm Im} \left[\frac{1}{M^2-m_h^2 + iM\Gm_h(M) } \right] \Pi(q,M) 
\ee
with $N_h=\displaystyle\int^{(m_h+2\Gm_h)^2}_{(m_h-2\Gm_h)^2}
dM^2\frac{1}{\pi} {\rm Im}\left[\frac{1}{M^2-m_h^2 + iM\Gm_h(M)} \right]$ and $\Gm_h=\Gm_h(m_h)$. Here $h\equiv\rho$ so that
\begin{equation}
\Gamma_{\rho}(M)=\Gamma_{\rho\rightarrow\pi\pi}(M) = \left[\frac{g_{\rho\pi\pi}^2}{48\pi M^3}\right]\left[ M^2-4m_\pi^2 \right]\lambda^{\frac{1}{2}}\left(M^2,m_\pi^2,m_\pi^2\right)
\end{equation}
where $\lambda(x,y,z)=x^2+y^2+z^2-2(xy+yz+zx)$. This is obtained using $\mathcal{L}_{\rho\pi\pi}=g_{\rho\pi\pi}\vec\rho_\mu\cdot(\vec\pi\times\del^\mu\vec\pi)$ with $g_{\rho\pi\pi}=6.05$.

For baryons ($R$) with non-trivial decay width in the loops we use
\be
\Pi(q,m_R)= \frac{1}{N_R}\int^{m_R+2\Gm_R}_{m_R-2\Gm_R}dM\frac{1}{\pi} 
{\rm Im} \left[\frac{1}{M-m_R + \frac{i}{2}\Gm_R(M)} \right] \Pi(q,M) 
\ee
with $N_R=\displaystyle\int^{m_R+2\Gm_R}_{m_R-2\Gm_R}dM\frac{1}{\pi} {\rm Im} 
\left[\frac{1}{M-m_R + \frac{i}{2}\Gm_R(M)} \right]$ and $\Gm_R=\Gm_R(m_R)$. In this case $R\equiv\De$ for which the decay formula is given by eq.~(\ref{deltadecay}) in Section III.

\subsection{Numerical results}

\begin{figure}
\centering
\includegraphics[scale=0.3,angle=-90]{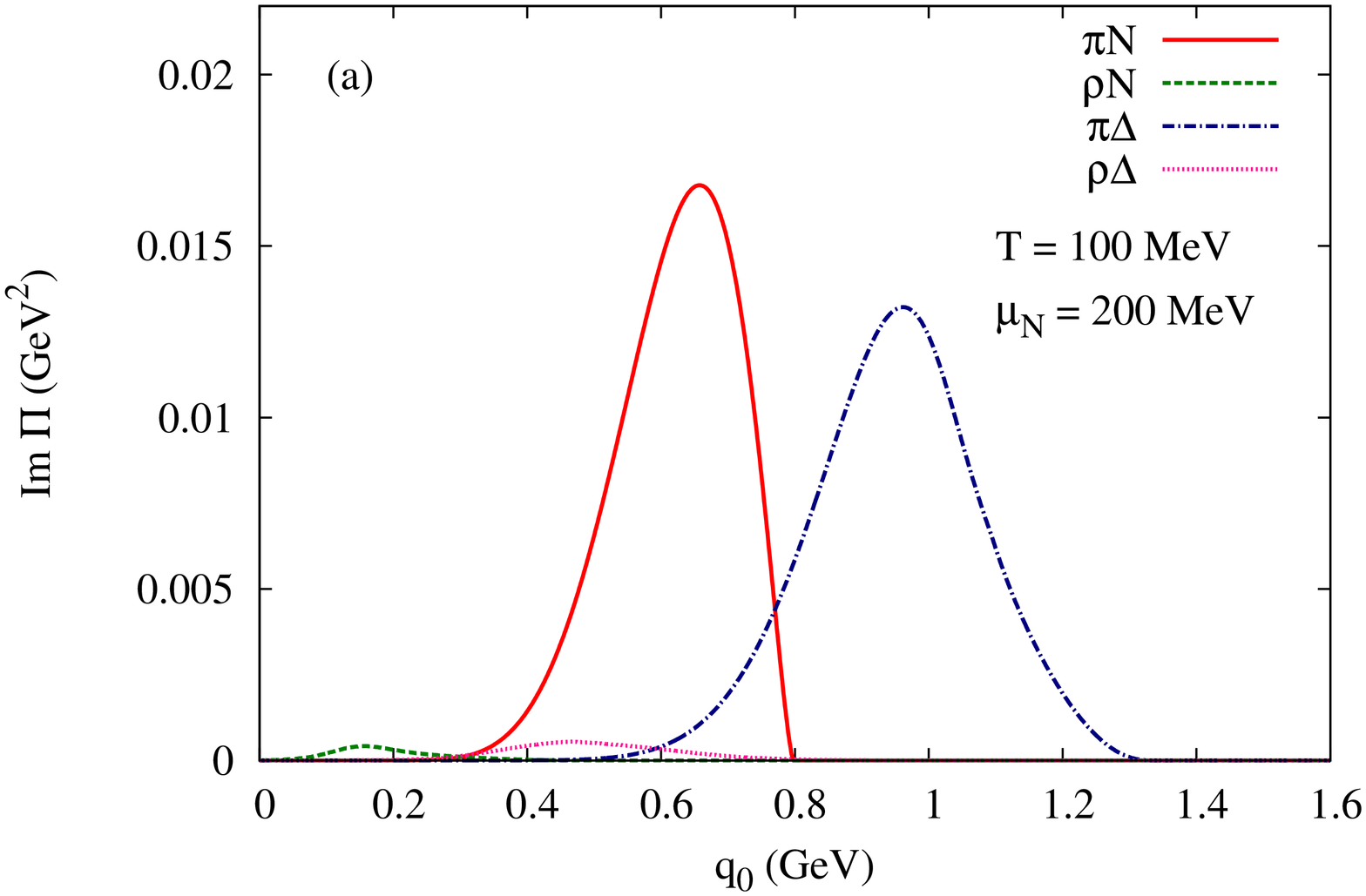}
\includegraphics[scale=0.3,angle=-90]{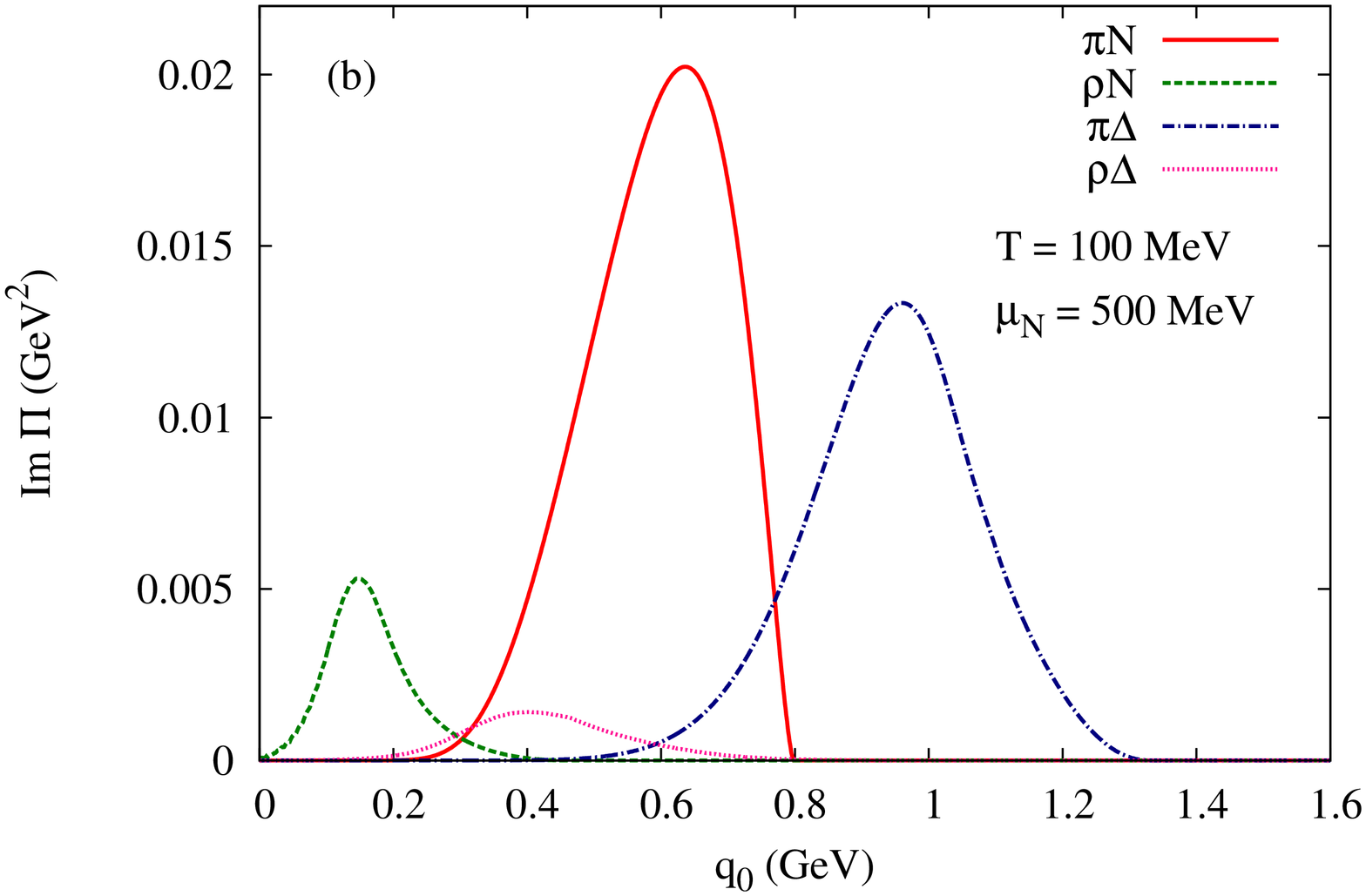}
\includegraphics[scale=0.3,angle=-90]{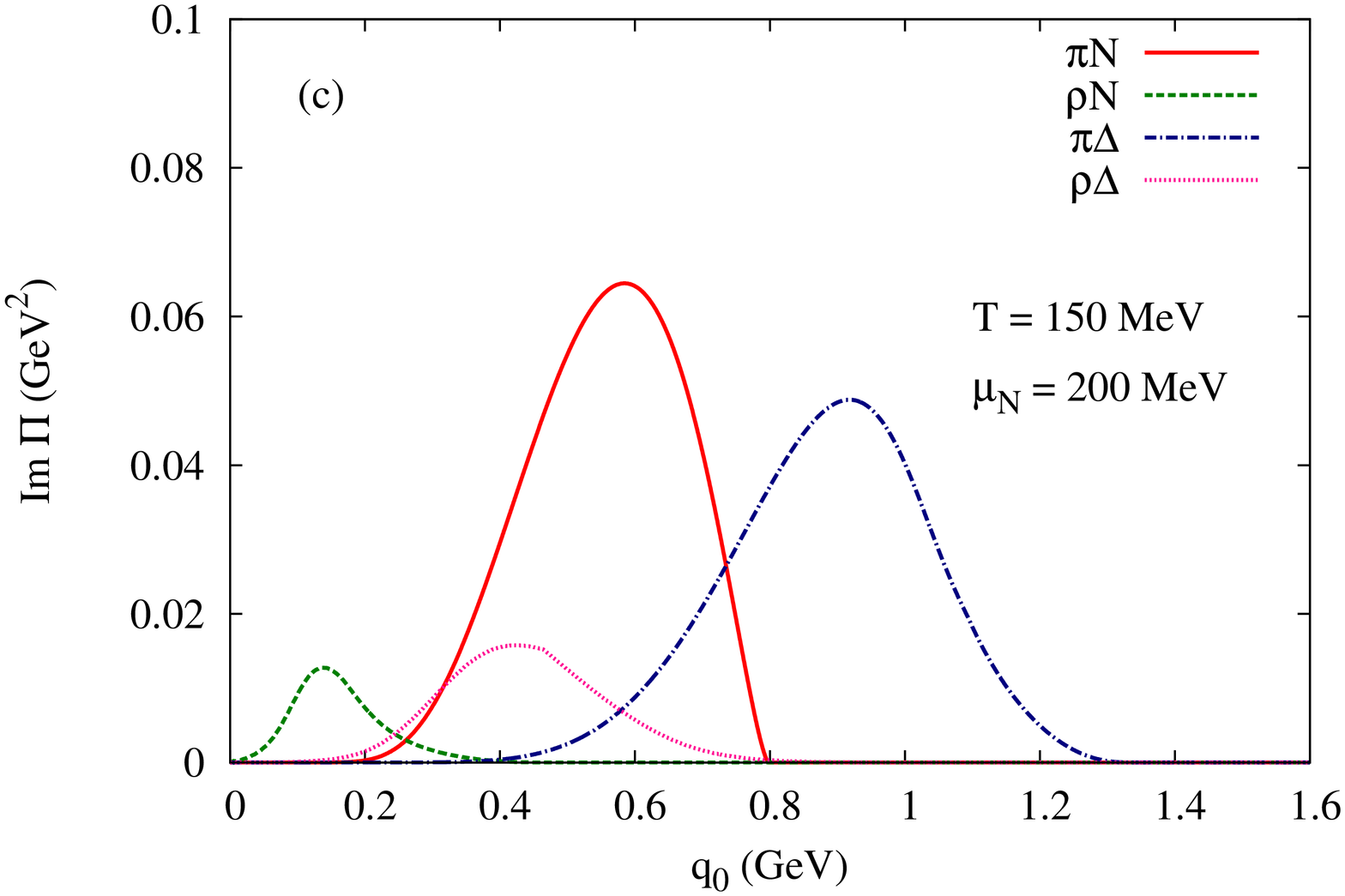}
\includegraphics[scale=0.3,angle=-90]{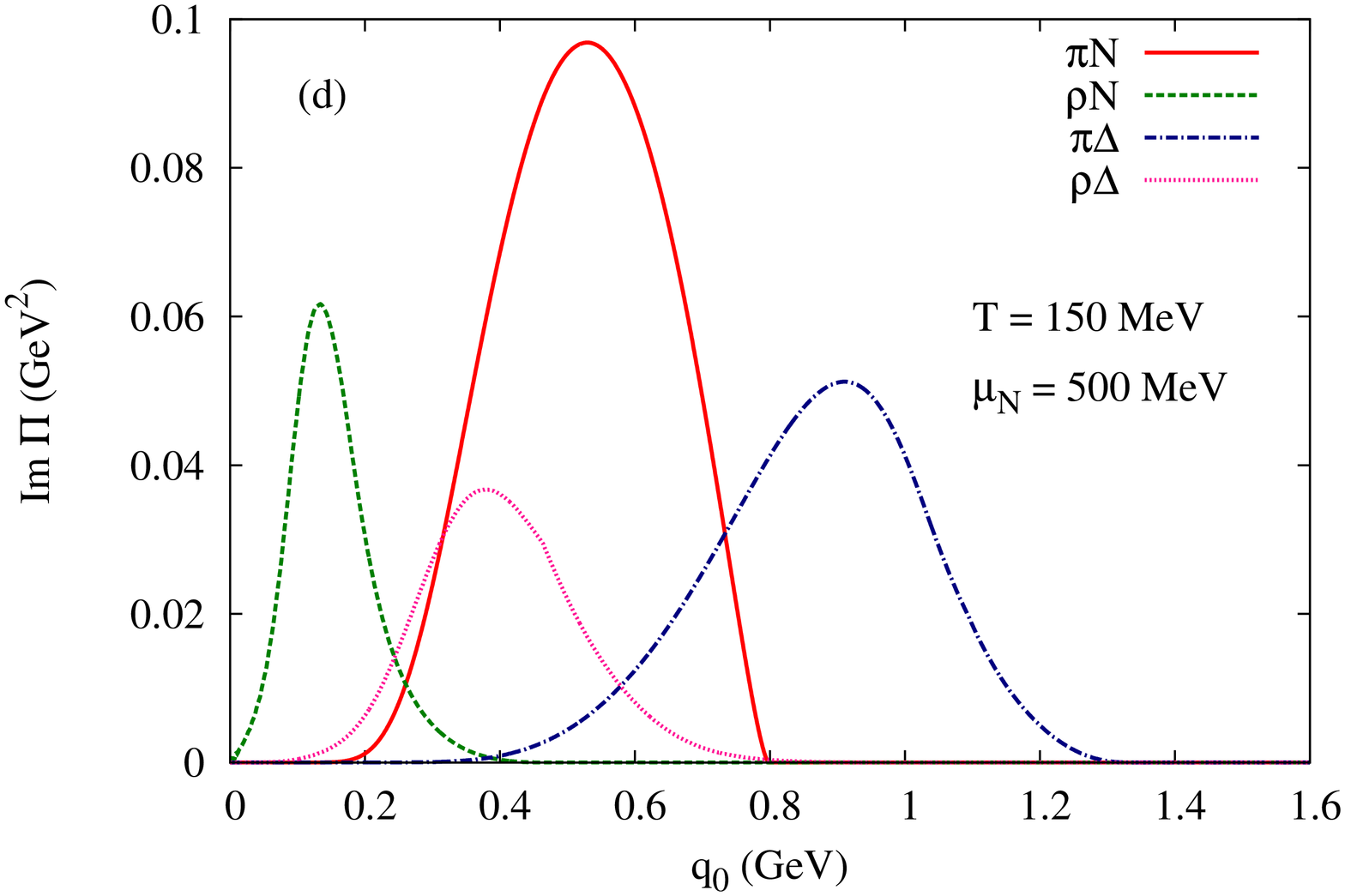}
\caption{The Landau-cut contribution to the imaginary part of the self energy function for different $T$ and $\mu_N$}
\label{fig:imag_Landau}
\end{figure}

Let us begin with the results for the $\De$ self-energy in the medium. We show numerical results for the spin-averaged real and imaginary parts of the function $\Pi$ given by~\cite{Wehrberger,Kim}
\begin{equation}
{\Pi}=\frac{1}{4}\sum_{s_\De}\bar{\Psi}_\mu\bar{\Pi}^{\mu\nu}\Psi_\nu
\label{spin-average}
\end{equation}
where $\bar{\Psi}_\mu$ denote Rarita-Schwinger spinors. The factors $N^\mn$ given in eqs.~(\ref{nmunu}) then go over to $\frac{1}{4}\text{Tr}[N^\mn\Sg_\mn]$. In fig.~\ref{fig:imag_Landau} we plot the Landau cut contribution to the imaginary part coming from the different loop graphs at $T=100$ MeV for $\mu_N=200$ MeV  and $\mu_N=500$ MeV in panels (a) and (b) respectively. The corresponding results for $T=150$ MeV are shown in panels (c) and (d). As discussed in the last section, it follows that for $\vec q=0$ the Landau cut extends up to the difference of masses of the baryon and the meson in the loop graph in the stable limit. These contributions at lower values of $q_0$ are a result of scattering processes in the thermal medium and are absent in vacuum. Comparing fig.~\ref{fig:imag_Landau}(a) with (b) (and (c) with (d)) we see that the $\rho N$ and $\rho\De$ loops start contributing to the imaginary part only at larger baryon densities.

\begin{figure}
\centering
\includegraphics[scale=0.4,angle=-90]{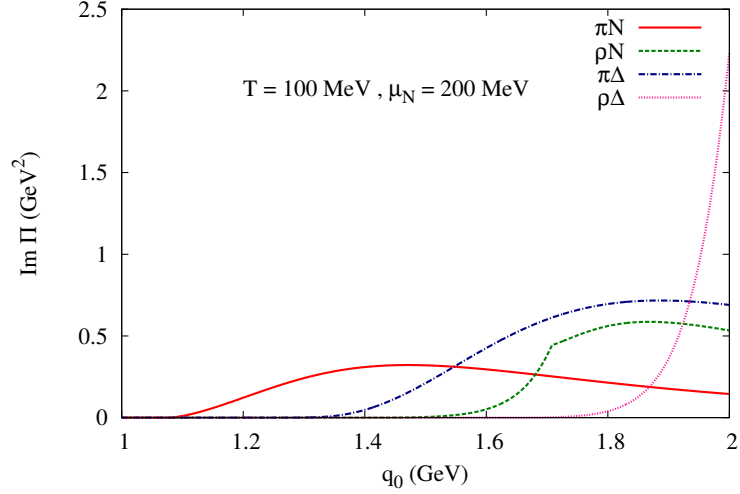}
%~~~~\includegraphics[scale=0.4,angle=-90]{impiu_150_200.eps}
\caption{The unitary cut contribution to the imaginary part of the $\De$ self-energy at $T$=100 MeV.}
\label{fig:imag_U}
\end{figure}

The unitary cut contributions to the imaginary part are much larger in magnitude than the ones coming from the Landau cut. The thresholds lie at higher energies, for stable particles starting from the sum of the nominal masses of the particles in the loop graph. Shown in 
fig.~\ref{fig:imag_U} are the unitary cut contributions from the four loops at $T=100$ MeV. We observe the sequential opening up of heavier decay channels. These are the same as in vacuum but are now weighted by the Pauli blocking and Bose enhancement factors in the final state. The form factor suppresses the usual monotonous rise of these contributions at higher $q_0$.

\begin{figure}
\centering
\includegraphics[scale=0.3,angle=-90]{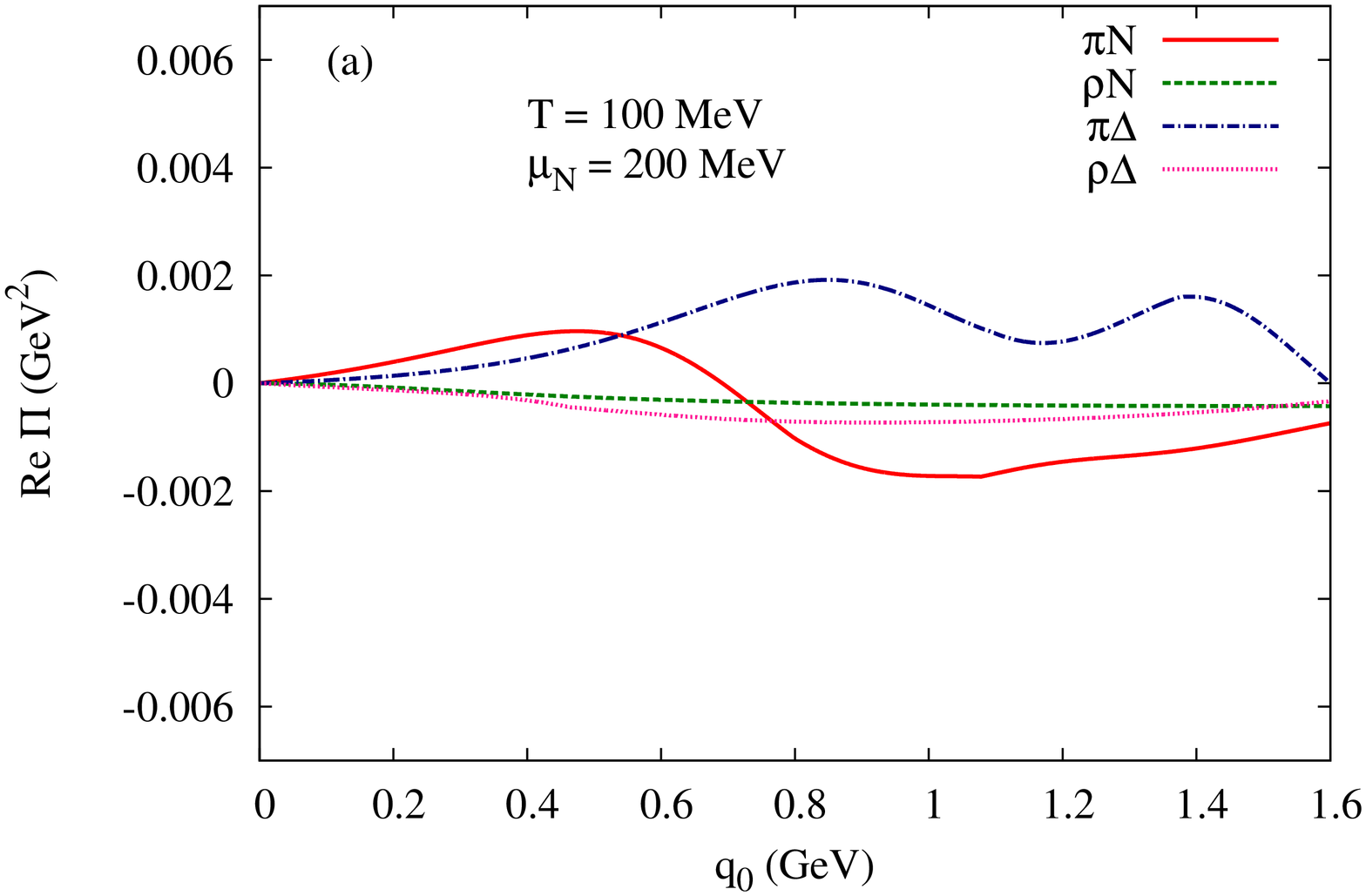}
\includegraphics[scale=0.3,angle=-90]{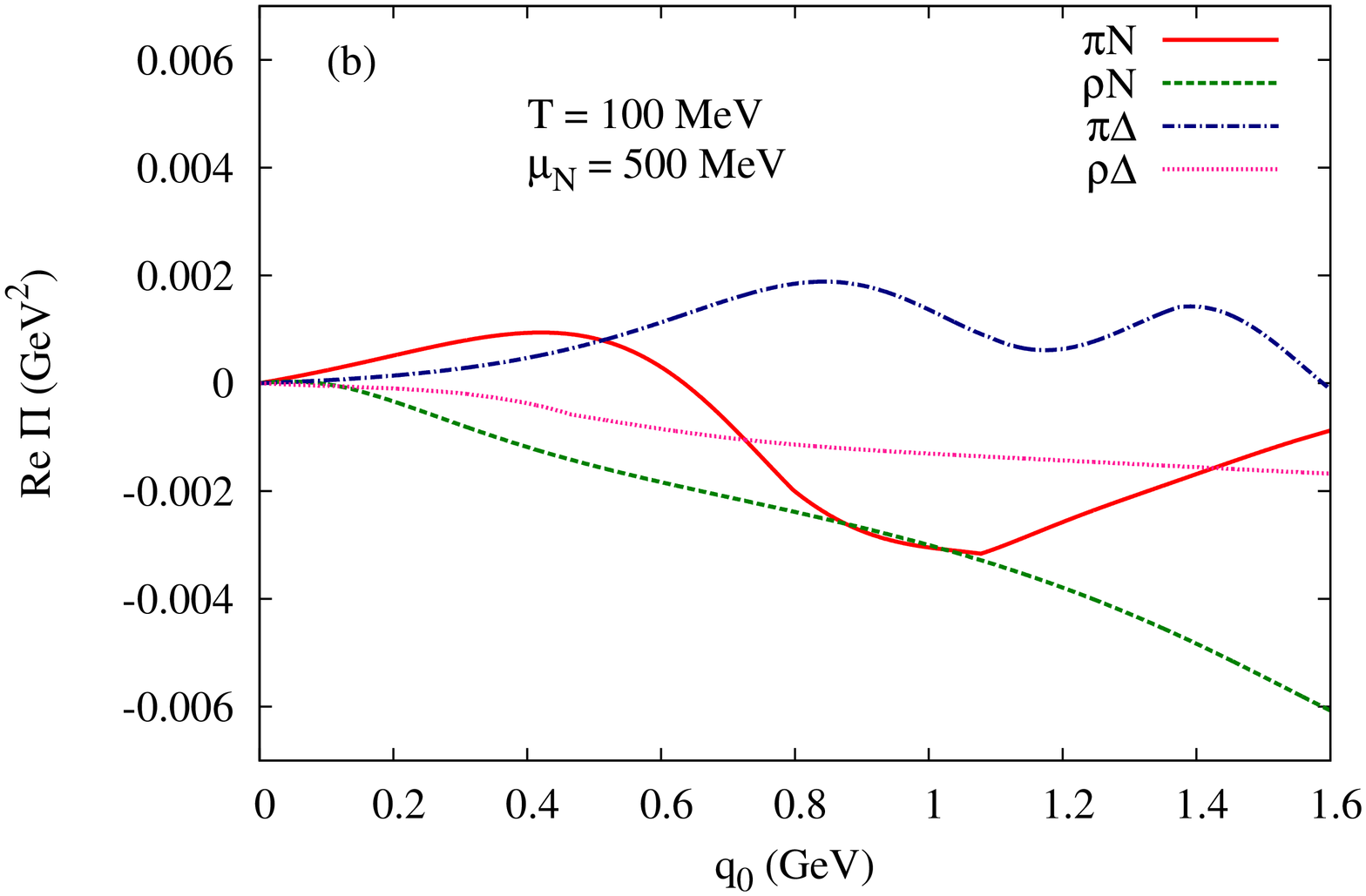}
\includegraphics[scale=0.3,angle=-90]{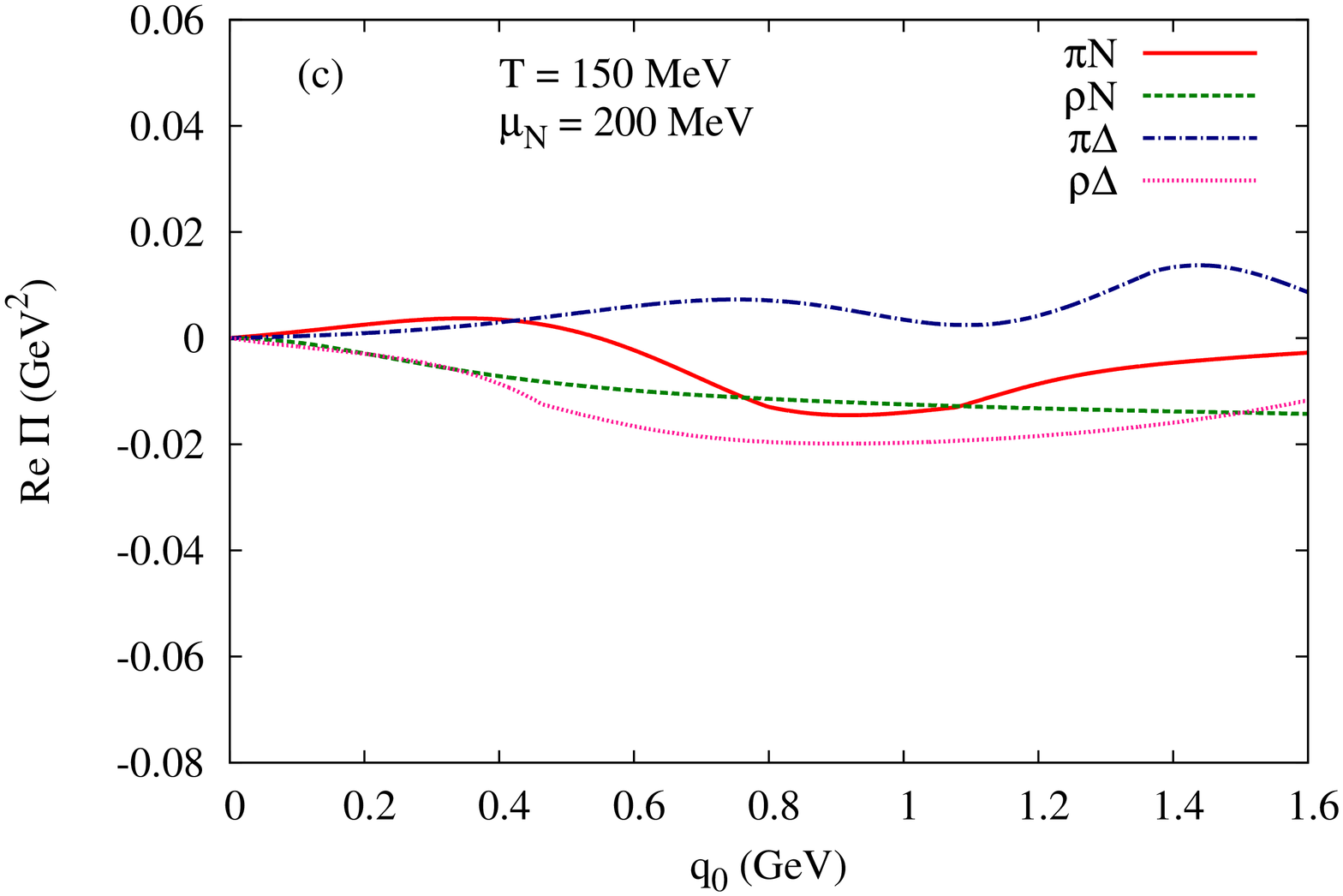}
\includegraphics[scale=0.3,angle=-90]{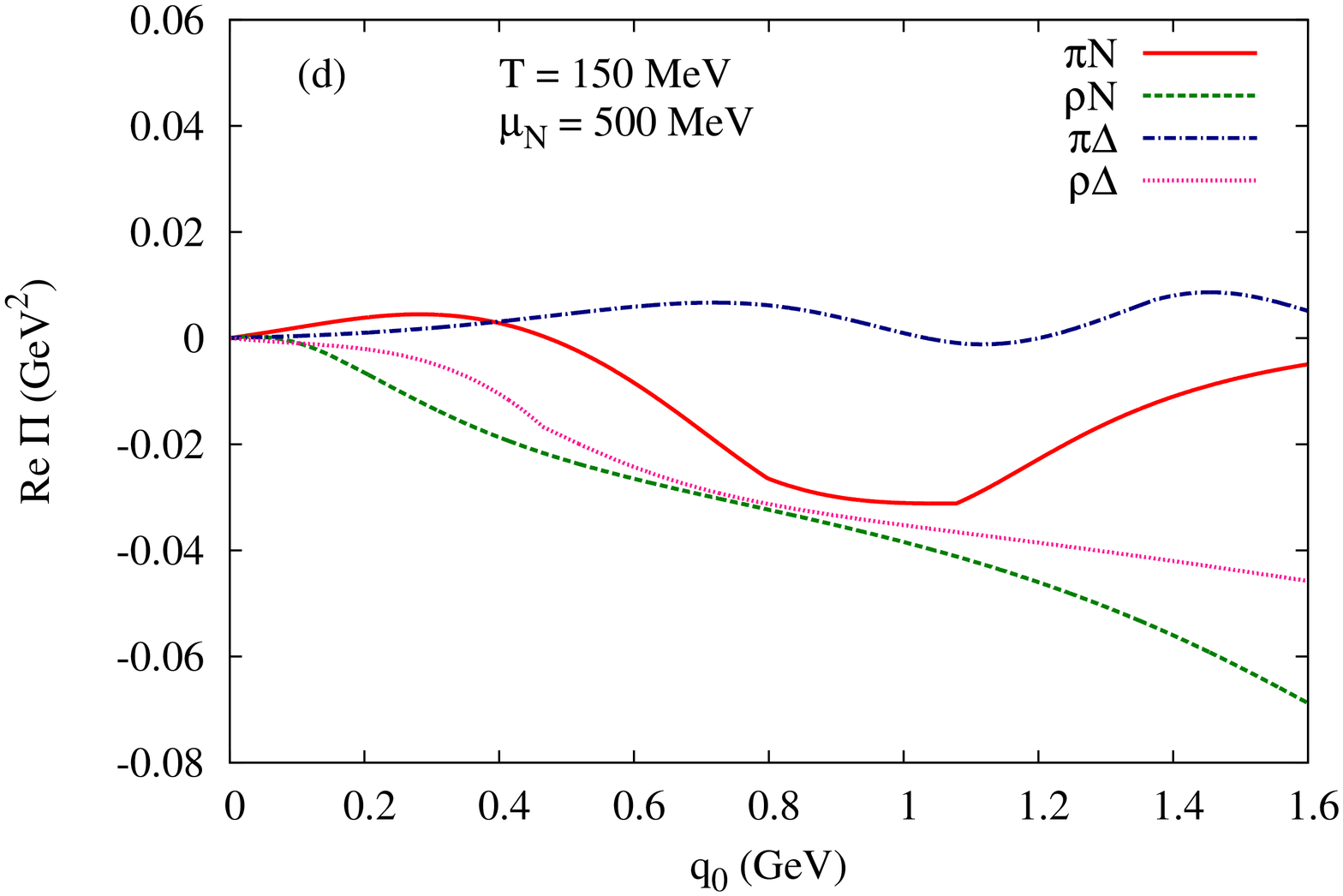}
\caption{The real part of the self energy function for different $T$ and $\mu_N$}
\label{fig:real}
\end{figure}

We now consider the thermal component of the real part of the self-energy consisting of principal value integrals. As seen in fig.~\ref{fig:real} the magnitudes are quite small compared to the imaginary parts and are not expected to contribute to a thermal shift in the pole position of the $\De$.

\begin{figure}
\centering
\includegraphics[scale=0.3,angle=-90]{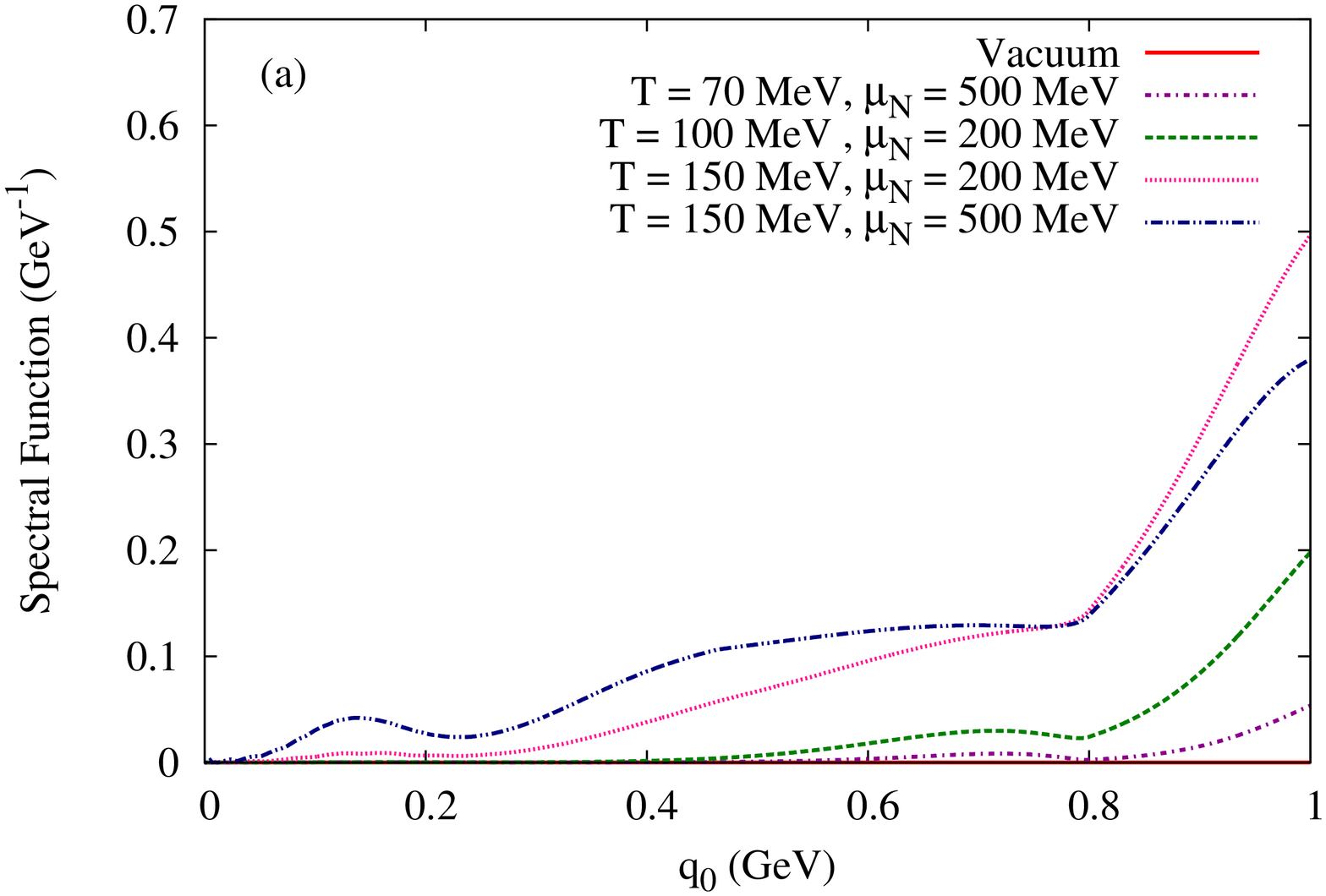}
\includegraphics[scale=0.3,angle=-90]{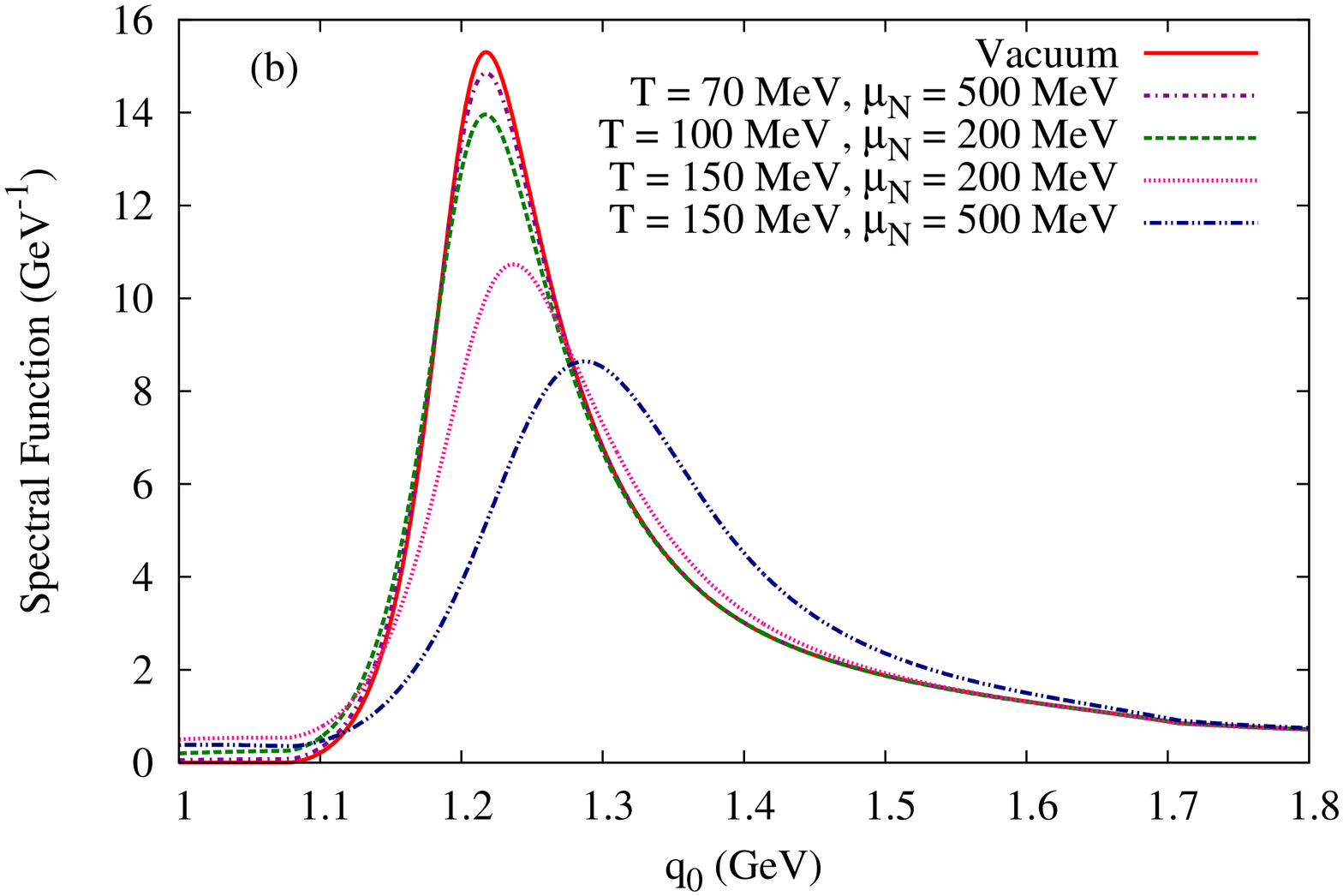}
\caption{The spectral function of the $\De$ for (a) $q_0$ up to 1 GeV and (b) $q_0$ from 1 to 1.8 GeV }
\label{fig:specfn}
\end{figure}

Having obtained the imaginary and real parts of the self-energy we now plot the spectral function which is the  imaginary part of the complete propagator~(\ref{dyson-scalar}). We do so
in two parts. The low $q_0$ region depicting the Landau cut contribution shown in panel (a) of fig.~\ref{fig:specfn} is purely a thermal contribution. The high $q_0$ region in the vicinity of the bare $\De$ mass consisting of the contributions from the unitary cuts is shown in panel (b) of fig.~\ref{fig:specfn}. The spectral density is seen to have a significant dependence on the  temperature and chemical potential of the medium. However, comparing the curves at $T=70$ and $T=150$ MeV, both for $\mu_N=500$ MeV, the effect of baryon density only shows up at higher temperatures and has its origin in the thermal distribution functions for the baryons. In general, we find a gradual suppression of the peak with increasing temperature and density owing to the larger imaginary parts in the denominator of the in-medium propagator. As seen from eq.~(\ref{im1}), the increase in the imaginary part comes from two factors: (i) the Bose enhancement factor for the pions and rho mesons in the final state in the first term which is the unitary cut contribution and (ii) the Landau cut contribution coming from scattering of the mesons off the propagating $\De$ as given by the fourth term. The second and third terms do not contribute because of kinematic reasons. 

\begin{figure}
\centering
\includegraphics[scale=0.3,angle=-90]{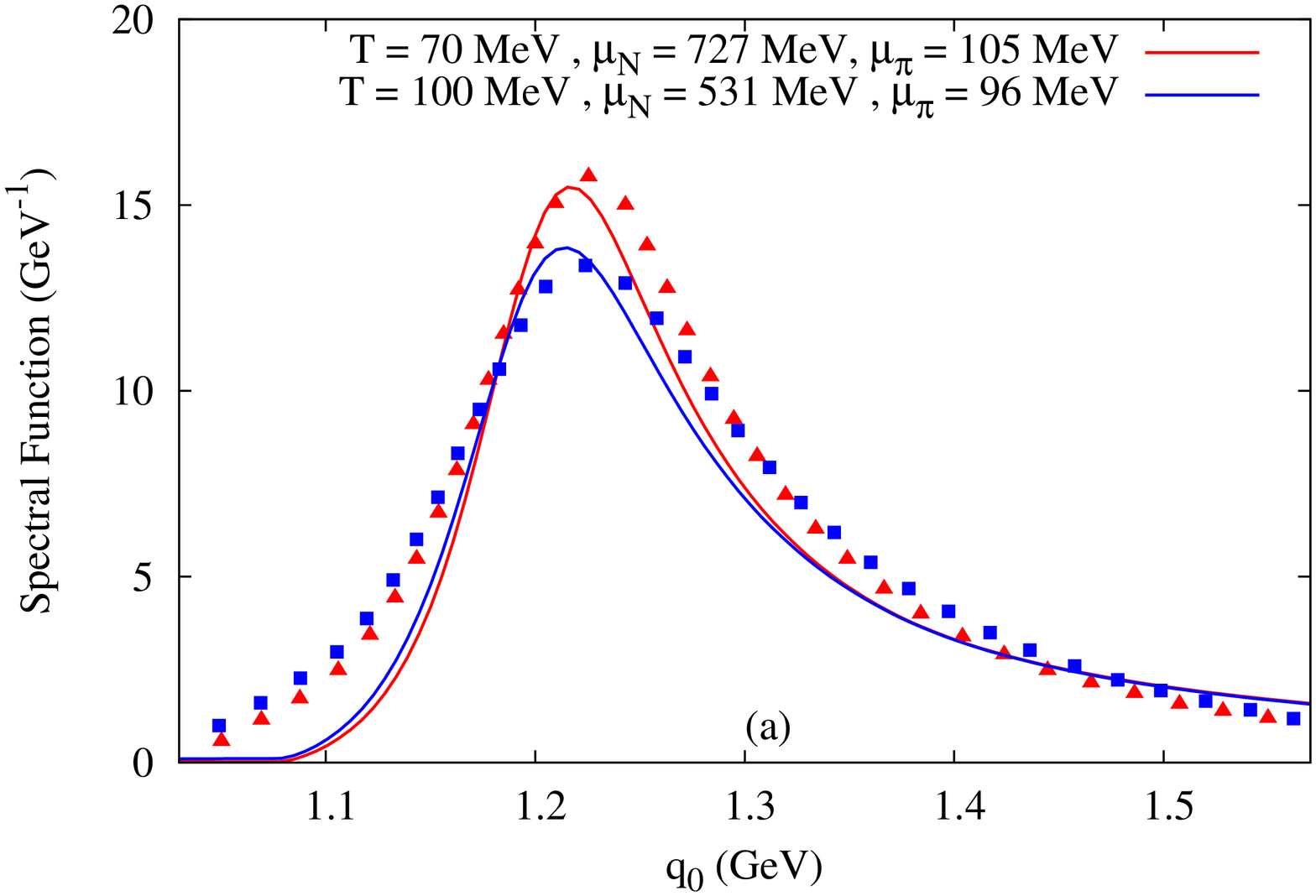}
\includegraphics[scale=0.3,angle=-90]{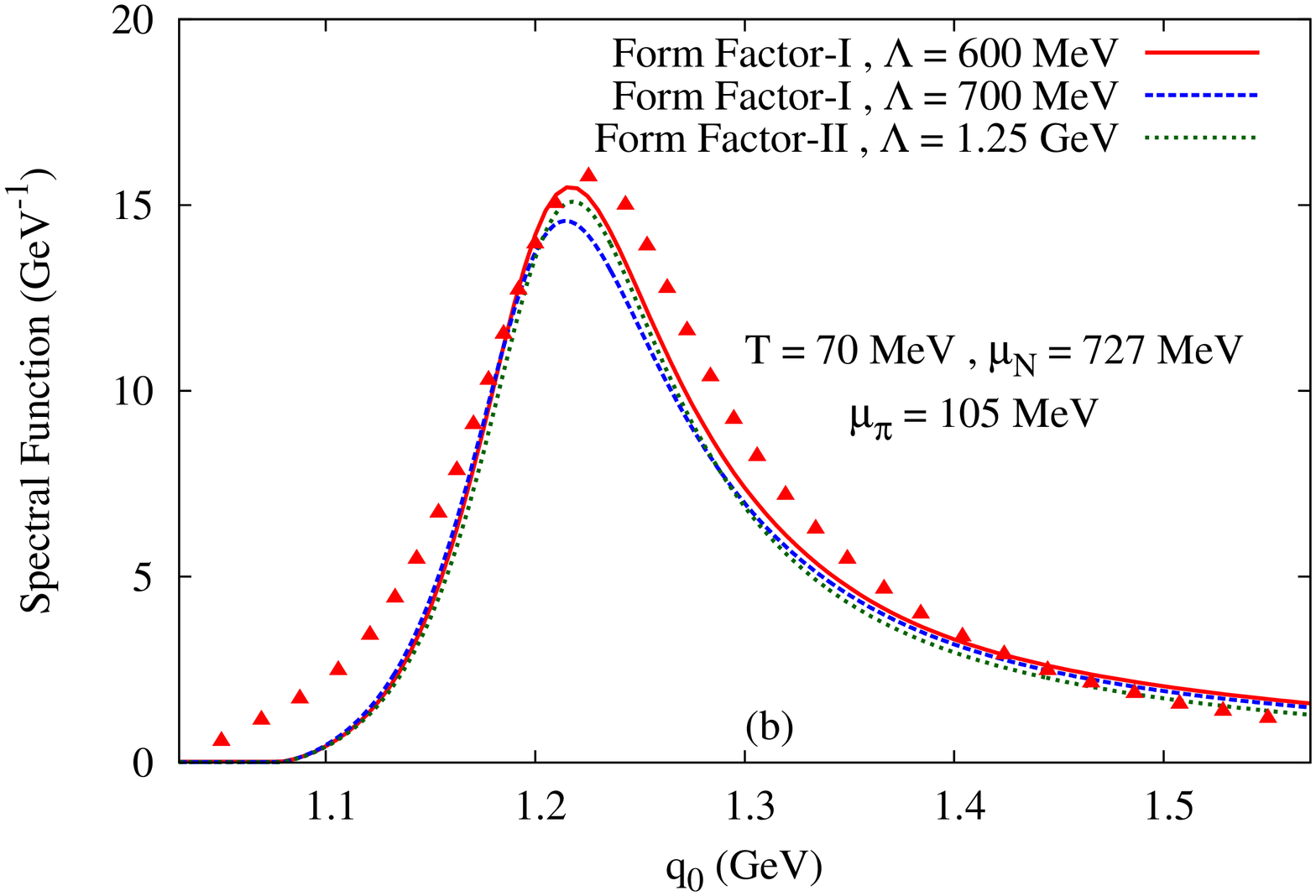}
\caption{The $\De$ spectral function as a function of $q_0$: (a) in comparison with~\cite{vanHees} and (b) for different form factors. The symbols correspond to the results of Ref.~\cite{vanHees}}
\label{fig:compare}
\end{figure}

In fig.~\ref{fig:compare}(a) we have made a comparison of the $\De$ spectral function obtained in our approach with that of~\cite{vanHees} for two sets of values of temperature and nucleon density representative of conditions likely to be achieved in relativistic heavy ion collisions at RHIC and in the CBM experiment. A reasonable agreement is observed between the present work and that of~\cite{vanHees} as depicted by the continuous lines and symbols respectively. In addition to differences in the Lagrangian and associated parameters used in the two approaches the disparities in the spectral functions in the two cases could arise due to contributions coming from higher order effects introduced through dressed nucleon and pion propagators in the $\De$ self-energy considered in~\cite{vanHees}  wherein vertex corrections were included through Migdal parameters in the pion propagator.

To take into account the finite size of the vertices a phenomenological hadronic form factor has been introduced. The details are provided in Appendix-B. The numerical results presented here correspond to a monopole type form factor, denoted by Form Factor I in fig.~\ref{fig:compare}(b) with $\Lm=600$ MeV which produces a good fit to the phase shift data and $\pi N$ cross section. We also plot the $\De$ spectral function using Form Factor I with $\Lm=700$ MeV and find a small reduction at the peak though the $\pi N$ cross section remains largely unchanged. This is because in the spectral function the square of the form factor appears multiplicatively but in the cross-section its effect is largely canceled at the $s$-channel pole position as seen from eq.~(\ref{modmsq}) in the next section. For a comparison we also plot the spectral function using an exponential form factor denoted by Form Factor II with $\Lm=1.25$ GeV (as used in~\cite{Korpa1} with $\Lm=0.97$ GeV). No appreciable difference is found with the one with $\Lm=$ 600 MeV. All the plots in this figure correspond to $T=70$ MeV, $\mu_N=727$ MeV and $\mu_\pi=105$ MeV. The symbols denote the results of~\cite{vanHees}.

At this point a few comments on the hadronic form factors at the vertices are in order. It is well known that in local field theory involving spin-3/2 fields the redundant degrees of freedom associated with unphysical spin-1/2 fields are eliminated through the Rarita-Schwinger constraints~\cite{Rarita}. The interacting case is more complex and suffers due to participation of the spurious spin-1/2 components. 
A coupling consistent with gauge invariance of the Rarita-Schwinger field was constructed in~\cite{Pascalutsa} preserving the correct number of degrees of freedom. However, unphysical behavior in the computed tree-level cross-section results if the reaction is cut off by standard hadronic form factors~\cite{Vrancx}. In this work we have used the conventional $\pi N\De$ vertex, which does suffer from a small presence of spin-1/2 components both in vacuum and in nuclear matter at saturation density~\cite{Korpa1}. As described above, the cut-off in the form factor used here was obtained by fitting the $\pi N$ cross-section. The spectral functions evaluated using this (conventional) vertex were found~\cite{Korpa3} to differ slightly with the ones calculated using the consistent coupling discussed above if the same form factor is applied in both the cases. It was further shown that this difference could be eliminated if the additional momentum factor stemming from the higher derivative nature of the consistent interaction~\cite{Pascalutsa} 
was compensated either by an additional form factor term or by adjusting the cut-off values of the original form factor.

\section{The $\pi$-N Cross Section}

\begin{figure}
\centering
\includegraphics[scale=0.3,angle=-90]{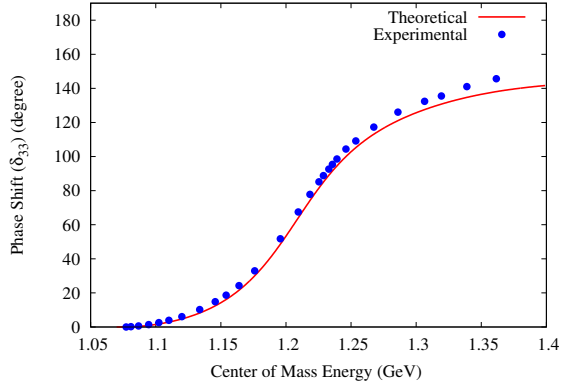}
\caption{The $\pi N$ phase shift $\de_{33}$ compared to data.}
\label{fig:phase}
\end{figure}

Having studied the spectral modification of the $\De$ in the medium we are now in a position to investigate how these changes affect the $\pi N$ cross-section. We aim to set up a dynamical framework wherein medium effects can be implemented using thermal field theoretic methods and which at the same time is normalized to the experimental data in vacuum. We consider the $\pi N\De$ interaction (\ref{Lagrangian_Pi_N_Delta}) and first check with the phase shift data~\cite{Koch} defining $\tan(\de_{33})=\frac{\text{Im}f}{\text{Re}f}$ with the partial wave amplitude given by
$f(E)\sim 1/[E^2-m_\De^2+im_\De\Gm_\De(E)]$. The  $\De\to \pi N$ decay width which follows from the imaginary part of 
eq.~(\ref{piN_vac}) is 
\be 
\Gm_\Delta(E)=\frac{1}{24\pi}\left(\frac{f_{\pi N\De}}{m_\pi}\right)^2 F^2(E)\frac{\vec p^{~3}}{E^2}[(E+m_N)^2-m_\pi^2]
\label{deltadecay}
\ee
where the c.m. momentum $\vec p^{\ 2}=[E^2-(m_N+m_\pi)^2][E^2-(m_N-m_\pi)^2]/4E^2$. As seen in fig.~\ref{fig:phase}, a reasonable agreement is obtained using $\Lambda=600$ MeV and $m_\De=1234$ MeV.

\begin{figure}
\centering
\includegraphics[scale=0.4,angle=-90]{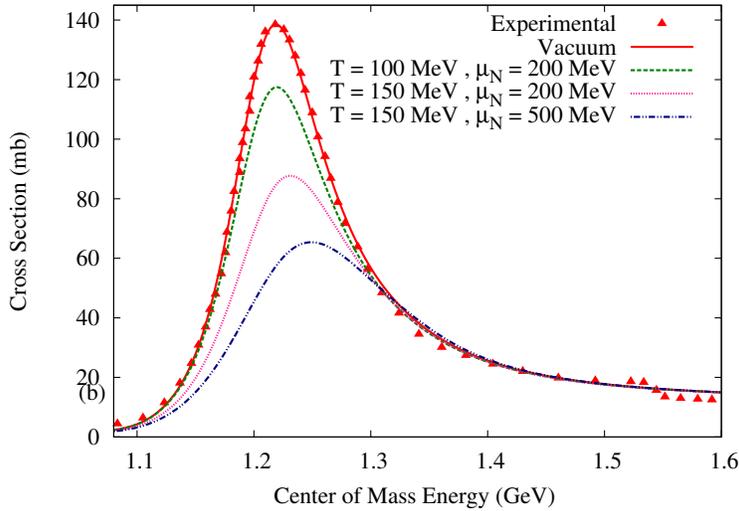}
\caption{The $\pi$-N elastic scattering cross section with medium effects.}
\label{fig:CrossSection}
\end{figure}

Next we evaluate the matrix elements for elastic $\pi N$ scattering in the isospin basis in which we replace the free vacuum $\De$ propagator by an effective one containing the vacuum self energy due to the loop diagrams mentioned above. 
Averaging over isospin, the squared invariant amplitude for the process $\pi(k)~N(p)\rightarrow~\pi(k')~N(p')$
is given by
\ba
\bar{|\mathcal{M}|^2}&=&\frac{\sum(2I+1)|\mathcal{M}_I|^2}{\sum(2I+1)}\nn\\
&=&\frac{1}{3}\left(\frac{f_{\pi N\Delta}}{m_\pi}\right)^4\left[ \frac{F^4(k,p)T_s}{\left|s-m_\Delta^2-{\Pi}\right|^2} + \frac{F^4(k,p')T_u}{\left(u-m_\Delta^2\right)^2}\right. \nn\\
&&\hskip 2.5cm\left.+ \frac{2F^2(k,p)F^2(k,p')T_m (s-m_\Delta^2-\text{Re}{\Pi})}{3(u-m_\Delta^2)\left|s-m_\Delta^2-{\Pi} \right|^2} \right]
\label{modmsq}
\ea
where $T_s$, $T_u$ and $T_m$ are given by
\begin{eqnarray}
T_s &=& \text{Tr}\left[(\cancel{p}'+m_N)D_s(\cancel{p}+m_N)\gamma^0D_s^\dagger\gamma^0 \right] \\
T_u &=& \text{Tr}\left[(\cancel{p}'+m_N)D_u(\cancel{p}+m_N)\gamma^0D_u^\dagger\gamma^0 \right] \\
T_m &=& \text{Tr}\left[(\cancel{p}'+m_N)D_s(\cancel{p}+m_N)\gamma^0D_u^\dagger\gamma^0 \right] 
\end{eqnarray}
in which
\begin{eqnarray}
D_s &=& k_\alpha k'_\beta\mathcal{O}^{\beta\nu}\Sg_\mn(q_s) \mathcal{O}^{\mu\alpha} \\
D_u &=& k'_\alpha k_\beta\mathcal{O}^{\beta\nu}\Sg_\mn(q_u) \mathcal{O}^{\mu\alpha}
\end{eqnarray}
where $\Sg_\mn$ is defined in (\ref{projectors}).

The  cross-section given by $\sigma(s)=\frac{1}{64\pi^2 s}\int\bar{|\mathcal{M}|^2}d\Omega$  turns out to be in good agreement with the isospin averaged total elastic cross-section given in~\cite{Prakash} (obtained using phase shift and inelasticity data from~\cite{Bareyre} and~\cite{Koch})% in the region of the $\De$ resonance 
up to about 1.5 GeV as seen from the solid curve in fig.~\ref{fig:CrossSection}. It is to be noted at this point that we have considered only $\De(1232)$ exchange in the evaluation of $\pi N$ elastic scattering amplitude with the aim of fixing the parameters (see Appendix B) and thus obtaining a baseline for estimating the effect of the modified $\De$ propagator on the cross-section. For a more general treatment it is necessary to consider the exchange of $N(938)$
as well as nearby resonances like the Roper(1440), $\De(1600)$ etc. in the evaluation of the scattering amplitudes. In such a  case, however, it could be quite challenging to obtain a satisfactory agreement with the experimental data especially in the region beyond the $\De$ peak.

Having thus normalized the framework with the experimental data we now turn on the medium effects. We replace the vacuum self energy in the above expressions by the in-medium ones evaluated in the real-time formalism described above. A significant suppression of the peak with increasing temperature is obtained owing to the increase in the imaginary part due to reasons explained earlier. The small upward shift at higher baryon densities
comes from the small positive contribution of the real part of the self-energy. As seen in fig.~\ref{fig:real}, there are substantial
cancellations between the contributions from various loops depending essentially on the attractive or repulsive nature of the effective interactions considered.

\section{The shear viscosity of a $\pi-N$ gas}

Transport coefficients can be obtained in (a) the kinetic theory approach using the transport equation and (b) the diagrammatic approach using Kubo formulae which relates them to retarded two-point functions. The latter was used in~\cite{Sabya} to obtain the shear viscosity of a pion gas. However, for our present purpose which is to highlight the effect of the in-medium $\pi N$ cross-section on the shear viscosity, the kinetic theory approach is more suited (see e.g~\cite{Mitra1,Mitra2}). The transport equation describing the evolution of the phase space density of pions and nucleons in a hadronic gas mixture slightly away from local equilibrium is given by
\be 
\frac{\del f_n}{\del t}+\vec v_n\cdot \vec\grad f_n=C[f_n]~~~~~~(n=\pi,N)
\label{transeq}
\ee
where $\vec v_n=\vec p/E_n$ is the particle velocity and $C[f_n]$ is the collision integral. The distribution function for such a system assumes a form $f_n=f_n^0+\delta f_n$ where the local equilibrium distribution function is given by $f_n^0=[\exp(p_n\cdot u-\mu_n)/T\pm 1]^{-1}$, the plus and minus signs correspond to nucleons and pions respectively and $\delta f_n$ is the deviation function. $T$, $u_\mu$ and $\mu_n$ denote the local temperature, fluid velocity and chemical potentials. Assuming all except the $n^{th}$ particle to be in equilibrium the collision integral simplifies to~\cite{Gavin}
\be 
C[f_n]\simeq -\frac{(f_n-f_n^0)}{\tau_n}=-\frac{\delta f_n}{\tau_n}
\label{cfk}
\ee
where $\tau_n$ is the relaxation time which characterizes the rate of change of the distribution function due to interaction with the species in the medium. For binary elastic collisions $p_n+p_l\to p'_n+p'_l$ it is~\cite{Prakash}
\be 
[\tau_n(p_n)]^{-1}=~\sum_{l=\pi, N}[\tau_{nl}(p_n)]^{-1}
\ee
with
\be
[\tau_{nl}(p_n)]^{-1}=\frac{g_l}{1+\delta _{nl}}\frac{{\rm csh}(\epsilon_{n}/2)}{E_n}\int d\omega _{l}d\omega_{n}^{'}d\omega _{l}^{'}W_{nl}
\label{taunl}
\ee
where $d\omega_k=d^3p_{k}/(2\pi)^3E_k[2~{\rm csh}(\epsilon_{k}/2)]$ , $\epsilon_k=(E_{k}-\mu_{k})/T$ and the function
${\rm csh}(x_k)=\cosh(x_k)(\sinh(x_k))$ if $k$ represents a fermion (boson). The dynamical input which goes into the determination of the distribution function appears in the interaction rate 
$W_{nl}=\frac{s}{2}\ \frac{d\sigma_{nl}}{d\Omega}(2\pi)^6\delta^4(p_{n}+p_{l}-p'_{n}-p_{l}')$.

To extract the shear viscosity we turn to the energy-momentum tensor. For small gradients of the local fluid velocity the
 shear dissipative part is well-known to be~\cite{Gavin}
\be 
T^{ij}_{shear}=-\eta\left(\frac{\del u^i}{\del x^j}+\frac{\del u^j}{\del x^i}-\frac{2}{3}\vec \grad\cdot\vec u\delta^{ij}\right)~.
\label{tij}
\ee
Now, in terms of the  distribution function the correction to the $ij$ component of the stress-energy tensor is given by~\cite{Itakura}
\be 
T^{ij}_{diss}=\sum_{n=\pi,N}g_n\int\frac{d^3p_n}{(2\pi)^3E_n}p_n^i p_n^j\delta f_n~.
\label{tij_dist}
\ee
From (\ref{transeq}) and (\ref{cfk}) we get to lowest order
\ba
\delta f_n&=&-\tau_n\left[\frac{\del f^0_n}{\del t}+\vec v_n\cdot \vec\grad f^0_n\right]\nonumber\\
&=&\frac{\tau_n f^0_n}{2TE_n}(1\pm f_n^0)p_n^ip_n^j\left(\frac{\del u^i}{\del x^j}+\frac{\del u^j}{\del x^i}-\frac{2}{3}\vec \grad\cdot\vec u\delta^{ij}\right)
\ea
where in the last line we retained only the (traceless) part appropriate for shear viscosity. Putting this in (\ref{tij_dist}) and equating with (\ref{tij}) we obtain the shear viscosity of the pion-nucleon mixture,
\be 
 \eta=\frac{1}{15T}\sum_{n=\pi, N}\int \frac{d^3 p_{n}}{(2\pi)^3}\frac{\tau_n(p_n)}{E_n^2}|\vec{p_n}|^4 {f_{n}^{0}(1\pm f_{n}^{0})}~.
 \ee

\begin{figure}
\centering
\includegraphics[scale=0.3,angle=-90]{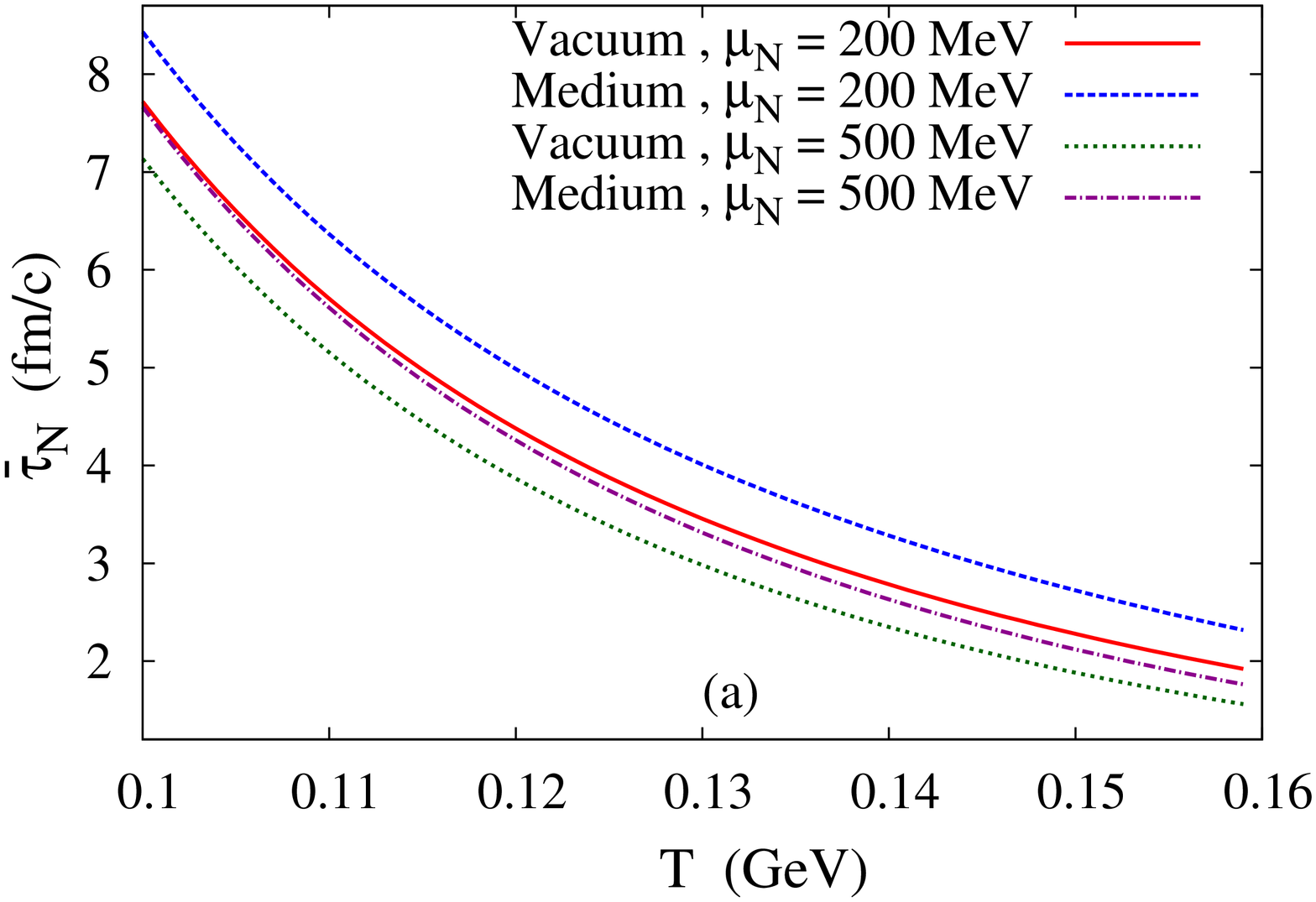}
\includegraphics[scale=0.3,angle=-90]{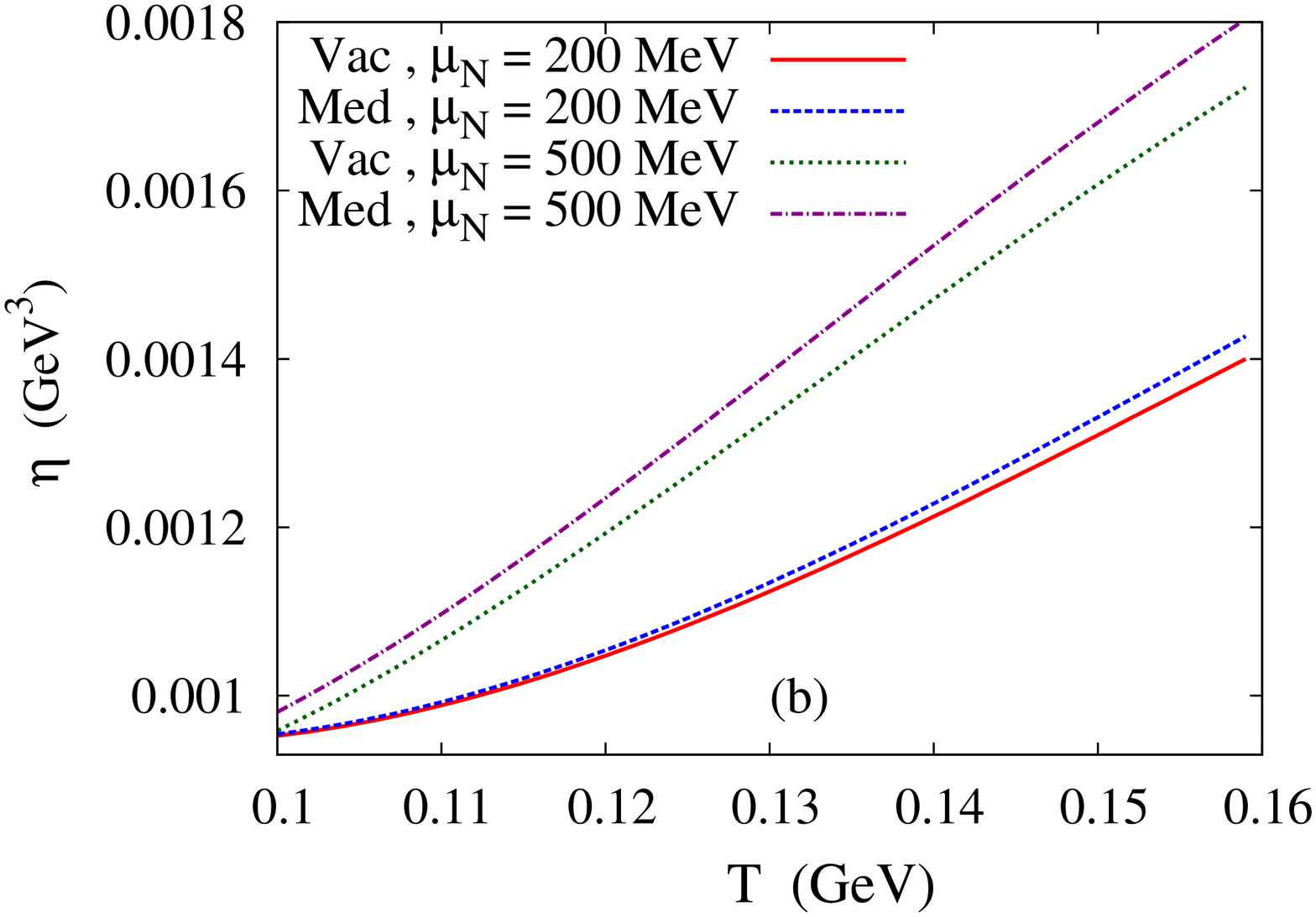}
\caption{The mean relaxation time of nucleons (a) and shear viscosity (b) of the $\pi N$ gas as a function of $T$ for two values of $\mu_N$. The legend 'Vacuum' indicates the use of $\pi N$ cross-section evaluated in vacuum in eq.~(\ref{taunl}).}
\label{fig:ShearViscosity}
\end{figure}

The shear viscosity is thus made up of contributions from the pion and nucleon components which are coupled through the momentum dependent relaxation times $\tau_N=[\tau_{N\pi}^{-1}+\tau_{N N}^{-1}]^{-1}$ and $\tau_\pi=[\tau_{\pi N}^{-1}+\tau_{\pi\pi}^{-1}]^{-1}$. The relative importance at a given value of $T$ and $\mu_N$ is a consequence of interplay between the phase space factors as well as scattering cross-sections. In order to focus on the effect of the in-medium $\pi N$ cross-section we take the $\pi\pi$ and $NN$ cross-sections in vacuum.

The mean relaxation time of the specie $i$ is defined in terms of the thermal average of the momentum dependent inverse relaxation time $\omega_i(k)=1/\tau_i(k)$.  With
  \be 
  \overline\omega_i(T,\mu)=\int d^3k \ \omega_i(k) f^0_i(k)/\int d^3k f^0_i(k)~,
  \ee
the mean relaxation time is given by $\overline \tau_i(T,\mu)=1/\overline\omega_i(T,\mu)$. We plot in fig.~\ref{fig:ShearViscosity}(a) the mean relaxation time of nucleons as a function of $T$ for two values of $\mu_N$ with and without medium effects. The features of the numerical results can be understood by realizing that the relaxation time for binary collision approximately goes as $\sim 1/n\sigma$ so that $\tau_N\sim[1-(\sigma_{NN}/\sigma_{N\pi})(n_N/n_\pi)]/\sigma_{N\pi}n_\pi$. Thus, the increase of density of the species, in this case pions, with $T$ plays the dominant role and accounts for the decreasing nature of the curves. It also follows that the relative increase of nucleon density for larger $\mu_N$ results in a relative decrease of the relaxation time. Again, since the pion density is considerably more than the nucleon density and $\sigma_{\pi N}$ is also much larger than $\sigma_{N N}$, the nucleon relaxation time is expected to be dominated by the pion component. Since $\sigma_{N\pi}$ is smaller compared to vacuum, as seen in fig.~\ref{fig:CrossSection}, the relaxation time is larger in the medium. 
Similar arguments hold also for the pion relaxation time. Its magnitude is decided by the (vacuum) $\pi\pi$ cross-section which being much larger, overshadows the medium dependence of the $\pi N$ cross-section. We thus do not show it separately. These features as well as the results with the vacuum cross-section are quite in agreement with~\cite{Prakash}.

In fig.~\ref{fig:ShearViscosity}(b) the shear viscosity is plotted as a function of $T$. As discussed above the behavior of the pion and nucleon components and their relative magnitude decides that of the mixture. For lower nucleon densities the pion component dominates the viscosity of the mixture. As the nucleon density increases there is a substantial increase in the nucleon component and a decrease in the pion component, the sum being more for $\mu_N=500$ MeV compared to 200 MeV. This feature is irrespective of the cross-section (vacuum or medium) and is due to the interplay between the relative abundance of the species and the magnitude of their interaction cross-section. The fact that the shear viscosity in the medium is more than that in vacuum can be understood in terms of the relaxation times which are larger basically due to the lower $\pi N$ cross-section in the medium. The increase in magnitude of the medium effect with temperature and nucleon density can be attributed to the phase space factors.

\section{Summary and Outlook}

In this work we have studied the spectral modification of the $\De$ baryon in the medium. The $\De$ self-energy was evaluated from one-loop graphs comprising of $\pi$, $\rho$, $N$ and $\De$ using the real-time formulation of thermal field theory. In addition to the contributions from decay processes occurring above thresholds which arise due to the usual thermally weighted unitary cut in the complex $q_0$ plane there are significant contributions coming from the Landau type discontinuities in the low $q_0$ region stemming from scattering processes leading to the absorption of $\De$ in the medium. The $\pi N$ cross-section  is then evaluated with the effective propagator of the $\De$ leading to a suppression at finite temperature and density with no significant shift of the peak position. This is expected to have non-trivial consequences on the mean free path and should consequently affect the thermalization rate of pions~\cite{Barz} and nucleons produced in heavy ion collisions. We finally make an estimate of the shear viscosity of a gas of pions and nucleons using the kinetic theory approach and observe an enhancement corresponding to the in-medium $\pi N$ cross-section which increases with temperature and nucleon density. When used as input in the hydrodynamic equations this is expected to have an observable consequence of the space-time evolution of the latter stages of heavy ion collisions.

\section{Appendix}
\subsection{Diagonalising the fermion propagator}

The procedure of diagonalization of the thermal matrices appearing in the real time formulation of thermal field theory only concerns the bosonic or fermionic nature of the field~\cite{Mallik}. The spin sum which appears in the numerator of the propagator can thus be factored out. The free thermal fermion propagator matrix is written as $\m{S}(p)=(\cancel p +m)\m{E}(p)$ where the matrix $\m{E}$ has components
\ba
&&E_{11}=-E^*_{22}=\De(p,m)-2\pi i\wt{N}^2_1\de(p^2-m^2)\nn\\
&&E_{12}=-2\pi ie^{\beta\mu/2}\wt{N}_1\wt{N}_2\de(p^2-m^2)\nn\\
&&E_{21}=2\pi ie^{-\beta\mu/2}\wt{N}_1\wt{N}_2\de(p^2-m^2)~.
\ea
The terms $\wt{N}_1$  and $\wt{N}_2$ containing thermal factors are given by
\ba 
\wt{N}_1(p_0)&& =\tht(p_0)\sqrt{\tnp}+\tht(-p_0)\sqrt{\tnm}\nn\\
\wt{N}_2(p_0)&& =\tht(p_0)\sqrt{1-\tnp}-\tht(-p_0)\sqrt{1-\tnm}\, 
\ea
where
\[\tilde{n}_\pm^p=\frac{1}{e^{\beta\left(\omega_p\mp\mu_p\right)}+1}~,~~~\omega_p=\sqrt{\vec{p}^2+m^2}~.\]

The matrix $\m{E}$ can be diagonalized as 
\be
\m{E}=\m{V}\left(\begin{array}{cc} \De & 0\\ 0 & -\De^*
\end{array}\right)\m{V}
\ee
where
\be
\m{V}=\left(\begin{array}{cc}\wt{N}_2 &
-\wt{N}_1e^{\beta\mu/2}\\
 \wt{N}_1e^{-\beta\mu/2} & \wt{N}_2
\end{array}\right)~.
\ee

It can be shown~\cite{Mallik} that the complete thermal propagator matrix $\m{S}'$ is also
diagonalized by $\m{V}$. From the Dyson equation (\ref{dyson-schwinger}) it then follows that the self energy matrix $\m{\Pi}$ is diagonalized by $\m{V}^{-1}$ resulting in eq.~(\ref{dyson-scalar}). The diagonal element $\ov\Pi$ is given by any one of the four components of $\m{\Pi}$ as given in eq.~(\ref{real-imag}).

\subsection{The $\De$ self-energy in vacuum : Lagrangian and parameters}
The $\De$ self energy in vacuum for the one loop diagrams 
shown in fig.~\ref{fig:Delta_SelfEnergy} are given by
\ba
\Pi^{\mu\nu}_{\pi N}&=& i\frac{f^2_{\pi N\Delta}}{m_\pi^2}\int\frac{d^4k}{(2\pi)^4}F^2(p,k)\mathcal{O}^{\nu\beta}k_\beta S^0(p)\mathcal{O}^{\alpha\mu}k_{\alpha}D^0(k)\label{piN_vac}\\
\Pi^{\mu\nu}_{\rho N}&=& i\frac{f^2_{\rho N\Delta}}{m_\rho^2}\int\frac{d^4k}{(2\pi)^4}F^2(p,k)\mathcal{O}^{\nu\eta}\gamma^5\gamma^\phi (g_{\beta\phi} k_\eta-g_{\beta\eta} k_\phi) S^0(p)  \gamma^5\gamma^\lambda (g_{\alpha\lambda} k_\sigma-g_{\alpha\sigma} k_\lambda)   \mathcal{O}^{\mu\sigma}D^{\alpha\beta}_0(k)\nn\\
&&\\
\Pi^{\mu\nu}_{\pi\Delta}&=& i\frac{f^2_{\pi\Delta\Delta}}{m_\pi^2}\int\frac{d^4k}{(2\pi)^4}F^2(p,k)\mathcal{O}^{\nu\chi}\gamma^5\gamma^\beta k_\beta\mathcal{O}^{\psi\sigma}g_{\chi\psi} S^0_{\lambda\sigma}(p)\mathcal{O}^{\lambda\eta}\gamma^5\gamma^\alpha k_{\alpha}\mathcal{O}^{\phi\mu}g_{\eta\phi}D^0(k)\\
\Pi^{\mu\nu}_{\rho\Delta}&=& if^2_{\rho\Delta\Delta}\int\frac{d^4k}{(2\pi)^4}F^2(p,k)\mathcal{O}^{\nu\chi} (\gamma^\beta+i\frac{\kappa_{\Delta\Delta\rho}}{2m_\Delta}\sigma^{\beta\epsilon} k_\epsilon)\mathcal{O}^{\psi\sigma}g_{\chi\psi} \nn\\
&& \hskip3cm S^{0}_{\lambda\sigma}(p)\mathcal{O}^{\lambda\eta}(\gamma^\alpha-i\frac{\kappa_{\Delta\Delta\rho}}{2m_\Delta}\sigma^{\alpha\delta} k_\delta)\mathcal{O}^{\phi\mu}g_{\eta\phi}D_{\alpha\beta}^{0}(k)
\ea
where $D^0(k)=\De(k,m_\pi)$ and $D^0_\mn(k)=A_\mn(k)\De(k,m_\rho)$ are the scalar and vector propagators in vacuum. The ones for the spin 1/2 and 3/2 fermions are given by $S^0(p)=(\cancel p +m)\De(p,m_N)$ and $S^0_\mn(p)=\Sigma_\mn(p)\De(p,m_\De)$ respectively.
The vertex factors come from the well-known interactions~\cite{Krehl} 
\begin{eqnarray}
\mathcal{L}_{\pi N\Delta} &=& \frac{f_{\pi N\Delta}}{m_\pi}\bar{\De}_\alpha{\cal O}^{\alpha\mu}\vec T^\dagger\partial_\mu\vec{\pi}\psi + H.c. \label{Lagrangian_Pi_N_Delta}\\
\mathcal{L}_{\rho N\Delta} &=&-i \frac{f_{\rho N\Delta}}{m_\rho}\bar{\De}_\alpha{\cal O}^{\alpha\mu}\gamma^5\gamma^\nu\vec T^\dagger \vec\rho_{\mn}\psi+ H.c. \\
\mathcal{L}_{\pi\Delta\Delta} &=& \frac{f_{\pi\Delta\Delta}}{m_\pi}\bar{\Delta}^\alpha{\cal O}_{\alpha\mu}\gamma^5\gamma^\nu\vec T\Delta^\mu\partial_\nu\vec{\pi}\\
\mathcal{L}_{\rho\Delta\Delta} &=&-f_{\rho\Delta\Delta}\bar{\Delta}^\beta{\cal O}_{\alpha\beta}\left[\gm^\mu-\frac{\kappa_{\rho\Delta\Delta}}{2m_\De}\sigma^\mn\partial_\nu\right]\vec\rho_\mu\vec T\De^\alpha
\end{eqnarray}
where~\cite{Krehl,Korpa1}, $f_{\pi N\Delta}=2.8$, $f_{\rho N\Delta}=16.03$, $f_{\pi\De\De}=1.78$, $f_{\rho\De\De}=7.67$ and $\kappa_{\rho\De\De}=6.1$.
In the above ${\cal O}_{\alpha\beta}=g_{\alpha\beta}-a\gm_\alpha\gm_\beta$ where the second term contributes only when the spin-3/2 field is off the mass shell. Thus the value of the coupling constants remain unchanged. 
At each vertex we consider the form factor~\cite{Ghosh2}
\be 
F(p,k)=\frac{\Lm^2}{\Lm^2+(\frac{p\cdot k}{m_p})^2-k^2}
\ee
in which $p$ and $k$ denote the momenta of the fermion and boson respectively. This form is denoted by Form Factor-I in fig.~\ref{fig:compare}(b). We determine the values of the  parameters $a$ and $\Lm$ by fitting the phase shift and vacuum cross-section. The height of the peak of the $\pi N$ cross-section is more sensitive to the value of $a$ and changes in $\Lm$ affect the tail at higher energies. The numerical results in this work have been generated with the values $a=0.002$ and $\Lm=600$ MeV. A reasonable fit it also obtained for $\Lm=700$ MeV. For comparison we also consider an exponential form factor~\cite{Korpa1} which we call Form Factor-II, given by
\be 
F(p)=\exp[-(p^2-(m_N+m_\pi)^2)/\Lm^2]
\ee
where $p$ is the momentum of the $\Delta$. In this case we obtain $\Lm=$ 1.25 GeV for the same value of $a$.

\end{document}